\documentclass[aps,pra,superscriptaddress,nofootinbib]{revtex4}
\pdfoutput=1
\usepackage[intlimits]{amsmath}
\usepackage{amsfonts}

\usepackage{comment}
\usepackage{subfigure}
\usepackage{enumitem}
\usepackage[usenames]{color}
\usepackage{braket}

\usepackage{graphicx}
\usepackage{epstopdf}

\advance\voffset by -1cm
\advance\hoffset by -0.5cm
\newcommand\be            {\begin{equation}}
\newcommand\ee            {\end{equation}}
\newcommand\bes           {\begin{subequations}}
\newcommand\esu           {\end{subequations}}

\newcommand{\bigx}[1]{\bBigg@{#1}}

\def\3pt#1#2#3{{\langle{#1}\vert{#2}\vert{#3}\rangle}}

\newcommand\doi[2]        {\href{http://dx.doi.org/#1}{#2}}

\newcommand{\EQ}{\begin{equation}}
\newcommand{\EN}{\end{equation}}
\usepackage{epsfig}
\usepackage{color}

\usepackage{amsmath}
\usepackage{graphicx}
\usepackage{amsfonts}
\usepackage{amssymb}
\usepackage{slashed}
\graphicspath{{Figures/}}

\begin{document}
\bibliographystyle{plainnat}

\title{{\Large {\bfseries Non Relativistic Limit of Integrable QFT \\ with fermionic excitations}}}
\author{Alvise Bastianello}
\affiliation{SISSA and INFN, Sezione di Trieste, via Bonomea 265, I-34136, 
Trieste, Italy}
\author{Andrea De Luca} 
\affiliation{The Rudolf Peierls Centre for Theoretical Physics, Oxford University, Oxford, OX1 3NP, United Kingdom}
\author{Giuseppe Mussardo}
\affiliation{SISSA and INFN, Sezione di Trieste, via Bonomea 265, I-34136, 
Trieste, Italy}

\begin{abstract}
\noindent
The aim of this paper is to investigate the non-relativistic limit of integrable quantum field theories with fermionic fields, such as the $O(N)$ Gross-Neveu model, the supersymmetric Sinh-Gordon and non-linear sigma models. The non-relativistic limit of these theories is implemented by a double scaling limit which consists of  sending the speed of light $c$ to infinity and rescaling at the same time the relevant coupling constant of the model in such a way to have finite energy excitations. For the general purpose of mapping the space of continuous non-relativistic integrable models, this paper completes and integrates the analysis done in Ref.\cite{first_paper} on the non-relativistic limit of purely bosonic theories.

\vspace{3mm}
\noindent
Pacs numbers: 

\end{abstract}
\maketitle

\maketitle

\section{Introduction}

In a previous paper by us \cite{first_paper} we were driven by the curiosity of understanding the striking unbalance which exists between the very large number of Integrable Quantum Field Theories (IQFT) and, in contrast, the very short list of Non-Relativistic Integrable Models\footnote{In Ref.\,\cite{first_paper} we were  concerned and we are also concerned here only with models which are local, galilean invariant and with field derivatives non higher that second order in their Hamiltonian density. Note that other NRIM such as the Calogero-Sutherland or the Haldane-Shastry models \cite{CSHS}, as well as all lattice models, do not belong to this class, since they are either non-local  or they are not continuum models, so they break relativistic invariance. } (NRIM). To analyze the origin of this richness/paucity difference between the two classes of models in Ref.\,\cite{first_paper} we pursued a study of the non-relativistic limit of a large set of IQFT (along the lines of the strategy adopted in earlier papers to study the non-relativistic limit of the Sinh-Gordon model \cite{KMT} and of the Thirring model \cite{dint}): for non-relativistic limit we mean not only sending the speed of limit $c$ to infinity but simultaneously tuning the relevant coupling constant of the model in order to keep the energy of its excitations finite. In Ref.\,\cite{first_paper} we were concerned with purely bosonic theories such as the Sinh-Gordon model \cite{AFZ1983,FMS1993}, the Bullough-Dodd model \cite{AFZ1983,DB1977,ZS1979,FMSBD,FLZZ} or, more generally, the Toda Field Theories \cite{Zamo1989,AFZ1983,BCDS1990,CM1989,ChM1990,M1992,Mbook,D1997,Oota,Delius} but also with the vectorial $O(N)$ non-linear sigma model \cite{ZZ}. 

The conclusion of the analysis presented in Ref.\,\cite{first_paper} was that, despite the vast diversity of the IQFT which were considered (they greatly differ for the spectrum of their massive excitations, the set of their coupling constants, exact S-matrices, etc.), all of them reduce in the non-relativistic limit  to the Lieb-Liniger model and closely-related generalizations thereof \footnote{A remarkable exception can be found in string theory \cite{reviewstring}, where the sigma model describing a superstring in an $\text{AdS}_5\times \text{S}^5$ background leads, in the NR limit and in the decompactification limit, to an integrable theory different from Lieb Liniger and generalization thereof. However, such a theory is no longer Galilean invariant and can be rather regarded as a lattice theory, the lattice space being the inverse of the string tension.} \cite{LL,yang-yang,yang,suth}. In more details, in the case of Toda Field theories, the outcome of taking their non-relativistic limit is a set of decoupled Lieb Liniger models while, for the $O(N)$ non-linear sigma model, a generalization of the Lieb Liniger model to many bosonic species, symmetrically coupled through a density-density interaction. 

In other words, in the case of bosonic IQFT, there are strong indications that the Lieb Liniger-like models may exhaust the list of possible NRIM with local interaction and galilean invariance.  Is there also a similar conclusion for IQFT involving fermionic fields? This is the question analysed in this paper.

The NR limit of the simplest fermion-based IQFT, namely the massive Thirring model, has already been considered in \cite{dint} and resulted in a generalized Lieb Liniger model with a pairwise local interaction involving derivatives of the fields (for generalizations, see also \cite{dint2}). Interestingly, the duality between the Thirring model and the purely bosonic Sine Gordon model \cite{coleman1975} is reflected also in their NR limits \cite{dint,dint2,CS1999}, being the NR limit of the Sine Gordon the attractive Lieb Liniger model \cite{dint,PKD2014}.
The purpose of the present investigation is to enlarge the landscape of the known NR limits, in particular we consider the supersymmetric Sinh-Gordon model (SShG) \cite{FGS1978,SM1986}, the $O(N)$ Gross-Neveu model (GN) \cite{ZZ,GN1978,ZZ1978} and the supersymmetric non-linear sigma model (SNLS) \cite{Witten-Shankar,Sc1990,A1994,A1991,Hollo,W1997}. Let's anticipate that the conclusions of 
this paper corroborate the same conclusions reached in \cite{first_paper}, i.e. adding fermionic fields into the game -- at least for the models here analyzed --  
does not break the tight bottleneck that projects IQFTs onto Lieb Liniger-like models and generalization thereof.

The paper is organised as follows: in Section \ref{Models} we introduce the models we are concerned with in the rest of the paper. In Section \ref{intsec} we provide a brief discussion on the basic features of integrable models. In Section \ref{SShGsec} we analyze the SShG model which,  among the three examples here considered, is the simplest and therefore it requires the smallest amount of technicalities, in particular its NR limit can be directly extracted from the Heisenberg equation of motion. In Section \ref{GNsec} we present the Gross-Neveu model and we implement its non-relativistic limit: in this case, additional work is needed and we must use the Feynman diagrams of the large $N$ expansion instead of the equation of motion.  Some technicalities related to the arguments of this section are included in the Appendix \ref{Gpropsec}. The last Section \ref{SNLS} analyses the supersymmetric non-linear sigma model, whose NR limit requires the same analysis used for the GN model. Finally, we gather our conclusions in Section \ref{conclusions}.

\section{IQFT's with fermionic excitations}\label{Models}

In this Section we briefly introduce the three fermionic models analysed in the rest of the paper. 

\subsection{The Supersymmetric Sinh-Gordon Model}

The SShG model is a IQFT based on a real bosonic field $\phi$ and on a Majorana spinor $\chi$, 
coupled in a supersymmetric way \cite{FGS1978,SM1986}
\be
\mathcal{S}_{\text{SShG}}=\int dx dt\; \left[\frac{1}{2}\partial_{\mu}\phi\partial^{\mu}\phi+\frac{1}{2}\bar{\chi}\left(i\slashed{\partial}-mc\cosh\left(g\frac{\phi}{2}\right)\right)\chi-\frac{m^2c^2}{g^2}\left(\cosh(g\phi)-1\right)\right]\, \, \, . \label{1}
\ee
If in the above action the fermionic field is ignored, one recovers the action of the Sinh Gordon model, so that SShG is the natural supersymmetric extension of the purely bosonic case. As already mentioned, in \cite{KMT} the proper NR limit of the Sinh Gordon model has been recognized to be the Lieb Liniger model and we 
will see that, consistently with this conclusion, in the NR limit the SShG model reduces to the supersymmetric extension of the Lieb Liniger model, namely the Yang-Gaudin model with a fermionic and a bosonic species \cite{LY1973,LY1974,ID2006,IDAnn2006}. This model consists in two species of NR particles coupled through a density-density interaction
\be
H^\text{NR}_\text{SShG}=\int dx \hspace{3pt}\left[\frac{\partial_{x}\varphi^{\dagger}\partial_{x}\varphi}{2m}+\frac{\partial_{x}\Psi^{\dagger}\partial_{x}\Psi}{2m}+\lambda\varphi^{\dagger}\varphi^{\dagger}\varphi\varphi+2\lambda \varphi^{\dagger}\varphi \Psi^{\dagger}\Psi \right]\,\,\, , \label{2}
\ee
where $\varphi$ and $\Psi$ are standard NR bosonic and fermionic fields respectively.  The details of this mapping, as well as the value of the coupling $\lambda$ resulting from the NR limit of the action (\ref{1}), are discussed in Section \ref{SShGsec}. 

\subsection{The $O(N)$ Gross Neveu Model}\label{introGN}
The GN model is a purely fermionic IQFT \cite{GN1978,ZZ1978,ZZ} which is based on $N$ majorana spinors $\chi_i(x,t)$, $i=1,2,\ldots N$ and with action 
\be
\mathcal{S}_{\text{GN}}=\int dxdt\; \left[\sum_{j=1}^N\frac{1}{2}\bar{\chi}_{j}i\slashed{\partial}\chi_j+\frac{\beta}{N}\left(\sum_{j=1}^N\bar{\chi}_{j}\chi_j\right)^2 \right]\, \, \, . \label{3}
\ee
In Section \ref{GNsec} we will show that, in the proper NR limit, the GN model reduces to a multicomponent Lieb Liniger model with fermionic particles \cite{yang,suth}
\be
H^\text{NR}_\text{GN}=\int dx \; \left[\sum_j \frac{\partial_x \Psi^\dagger_j\partial_x \Psi_j}{2m}+\lambda\sum_{j,j'}\Psi^\dagger_j \Psi^\dagger_{j'}\Psi_{j'}\Psi_{j} \right]\,\,\, ,\label{4}
\ee
where $\Psi_j$ are independent non relativistic fermionic fields. In the above summations, we discarded the upper bound of the indexes on purpose: in the NR limit, the GN model describes at once all the possible numbers of different species, as we will clarify in the main text where we will also provide the explicit value of the interaction (that turns out to be attractive, as expected from the fact that the GN model possesses bound states).

\subsection{The Supersymmetric Non-Linear Sigma Model}

The last model we consider in this paper is the supersymmetric non-linear sigma model (SNLS) \cite{W1997,Witten-Shankar}, 
that is an integrable supersymmetric version of the non-linear sigma model (NLS) whose NR limit has been extracted 
in \cite{first_paper}. Its action contains an equal number of real bosonic fields and majorana spinors
\be
\mathcal{S}_\text{SNLS}=\int dxdt\;\left[\sum_j \frac{1}{2}\partial_\mu\phi_j\partial^\mu\phi_j +\frac{1}{2}\sum_j\bar{\chi}_ji\slashed{\partial}\chi_j+\frac{\beta}{N}\left(\sum_j\bar{\chi}_j\chi_j\right)^2\right]\,\,\, ,\label{n5}
\ee
subjected to the constraints
\be
\sum_j\phi_j^2=\frac{N}{8}\beta^{-1},\hspace{3pc}\sum_j \phi_j\chi_j=0\,\,\, .\label{n6}
\ee
As it is clear from its action, the SNLS lies in between the Gross-Neveu model and the non-linear sigma model. 
Indeed, the non-relativistic limit discussed in Section \ref{SNLS} 
reflects this fact, with the corresponding Hamiltonian given by 
\be
H_\text{SNLS}^\text{NR}=\int dx\; \left[\sum_j\frac{\partial_x \varphi^\dagger_j\partial_x \varphi_j}{2m}
+ \sum_j\frac{\partial_x \Psi^\dagger_j\partial_x \Psi_j}{2m} +
\lambda\sum_{j,j'}\left(\varphi^\dagger_j\varphi^\dagger_{j'}\varphi_{j'}\varphi_j- \Psi^\dagger_j\Psi^\dagger_{j'}\Psi_{j'}\Psi_j+2 \varphi^\dagger_j\Psi^\dagger_{j'}\varphi_{j'}\Psi_j\right)\right]\,,\label{n7}
\ee
where $\varphi_i$ and $\Psi_i$ are bosonic and fermionic fields respectively. This model is a supersymmetric mixture equally suspended between the NR limit of the GN model and the NR limit of the non-linear sigma model found in \cite{first_paper}. The above non relativistic model belongs to a wider class of supersymmetric integrable models presented in \cite{FPZ1988}.

\section{Basic facts about integrability}
\label{intsec}

Integrability is nowadays a milestone in physics that continuously provides new exciting results: for example, in recent years we have witnessed 
its fundamental role in understanding -- in a very controllable way -- basic aspects of out-of-equilibrium physics as well as the subtilities of this topic, see for instance 
the papers (and references therein) collected in the special issue \cite{integrablequench}, moreover integrability is of fundamental importance also in String and Gauge theory (see \cite{reviewstring} for a review).
While in the classical realm the definition of integrability has been clearly stated by the Liouville theorem, i.e. a classical system with $N$ degrees of freedom is integrable if it possesses $N$ independent conserved quantities in involution, the definition is not equivalently crystal clear in the quantum realm. An extensive discussion about the rigorous definitions of integrability will eventually lead us far beyond the purposes of the present paper, but the interested reader can find in Ref.\,\cite{CM2011} a careful discussion about this issue. For what concerns us, we adopt the more pragmatic definition commonly used, alias a model is integrable if it possesses an infinite number of independent local conserved charges, i.e. conserved quantities that can be written as integrals over the space of a local operator (the density). The presence of such a huge number of conservation laws has immediate consequences on the scattering properties of the model, independently of being relativistic or not: in particular, many-body scattering events are always factorised in two-body elastic scattering processes  \cite{ZZ} (see also \cite{Mbook} and references therein). At this level, the most striking feature of integrability is 
displayed by relativistic theories. In fact, even though in principle relativistic invariance allows for particle production, the elastic scattering imposed by integrability forcefully prevents a net creation/destruction of particles: the number of particles remains always conserved. Indeed, checking the absence of particle production in the diagrammatic expansion of the scattering matrix is a quite efficient method to check the integrability of a given relativistic model \cite{D1997,first_paper}.
Apart from the conservation of the total number of particles (that could be regarded somewhat as a trivial statement in the non relativistic framework, where particle production is absent), integrability provides a powerful machinery to get exact, non perturbative, informations about a given model. In this respect it is important 
to mention three main approaches: integrable field theories \cite{Zamo1989, ZZ,Mbook, Smirnov}, the thermodynamic Bethe-Ansatz \cite{takahashi} and the algebraic Bethe-ansatz \cite{korepin}. 

The key quantity of the integrable models is the two-body scattering matrix 
\be
S_{j_1 j_2}^{j_3 j_4}(\theta_1-\theta_2)\,\,\, ,
\ee
where we conventionally refer the lower labels to the incoming particles and the upper ones to the outgoing particles. In the case of a relativistic scattering, 
for the Lorentz invariance of the process, the scattering matrix depends only on the difference of the rapidities which parameterize the relativistic dispersion relations of the particles. In the case of NR scattering, Galilean invariance imposes the scattering matrix to be instead a function of the relative velocity in the center of mass frame, the latter being a non-trivial function of the incoming momenta, when particles with different masses are considered. 
The factorisation of any scattering event in two-body processes 
implies some important constraints on the $S$ matrix, in order to ensure the consistency of the factorisation.
 
\begin{figure}[t]
\begin{center}
\includegraphics[scale=0.35]{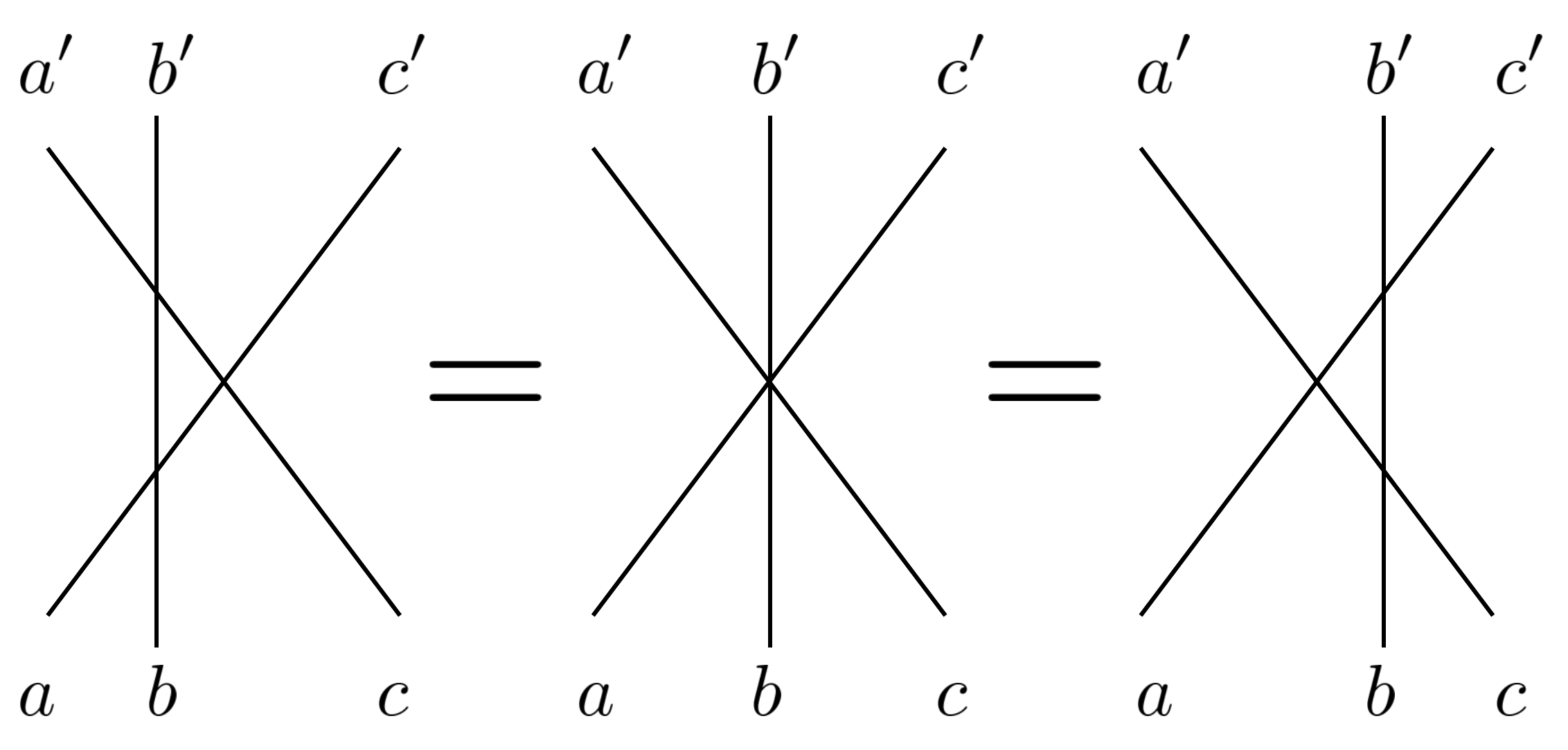}
\caption{\emph{Graphical representation of the Yang Baxter equations.}}\label{fig1}
\end{center}
\end{figure}

The first one is unitarity: this property is a direct consequence of the unitarity of the time evolution and must be respected by any physical scattering matrix, regardless the integrability. However, for the nature of the factorised scattering, the unitarity can be imposed at the level on the two-body scattering matrix
\be
S_{k_1 k_2}^{j_3 j_4}(-\theta)S^{k_1 k_2}_{j_1 j_2}(\theta)=\delta_{j_1}^{j_3}\delta_{j_2}^{j_4}\,\,\, , \label{6}
\ee
where we implicitly sum over the repeated indices. The second important constraint on the two-body scattering matrix are the Yang Baxter (YB) equations (summation over the repeated indexes is assumed)
\be
S_{jt}^{a'b'}(\theta_{23})S^{tc'}_{kc}(\theta_{23})S^{jk}_{ab}(\theta_{12})=S_{tk}^{b'c'}(\theta_{12})S_{aj}^{a't}(\theta_{13})S_{bc}^{jk}(\theta_{23})\label{7}
\ee
that guarantee the consistency of the two different ways in which a three body scattering event can be factorised in terms of two-body processes (see FIG. \ref{fig1}). Unitarity and YB naturally provide the consistency of the asymptotic state representation of the Hilbert space. 
Each state is represented as a set of ordered particles as
\be
\ket{A_{j_1}(\theta_1)A_{j_2}(\theta_2)A_{j_3}(\theta_3)...}\,\,\, .
\ee
Reshuffling the order of the rapidities can be interpreted as a scattering event: exchanging two nearby particles simply amounts to make them to scatter
\be
\ket{...A_{j_{n}}(\theta_n)A_{j_{n+1}}(\theta_{n+1})...}= S^{j_n j_{n+1}}_{k_{n} k_{n+1}}(\theta_n-\theta_{n+1})\ket{...A_{k_{n}}(\theta_{n+1})A_{k_{n+1}}(\theta_n)...}\,\,\, ,
\ee
where above we implicitly sum over the repeated indexes $k_n,k_{n+1}$.

The scattering matrix of integrable systems must satisfy both (\ref{6}) and (\ref{7}); moreover extra analytic conditions are required in the relativistic models, the most relevant being the \emph{crossing symmetry}. Actually these equations, beside being a check of integrability, can be even used constructively: in IQFT the \emph{S-matrix bootstrap program} \cite{Zamo1989,ZZ} consists of combining all the equations mentioned above together with a condition of analyticity 
of the amplitudes, with the aim to identify the exact $S-$matrixes of many different models. It must be said that for the bootstrap program implemented in a relativistic context a key role is played by the crossing symmetry, that is instead absent in non relativistic theories. However in NRIM we often do not need any bootstrap machinery, since the interaction is usually so simple that permits a calculation of the $S$-matrix from scratch. Nevertheless, it is still true that any solution to the unitarity and Yang Baxter equations is potentially a scattering matrix for a putative integrable model: while there exist suitable techniques to generate such solutions \cite{sklyanin1,kulish1,kulish2,sklyanin2,drinfeld,Jimbo3,Faddeev,Bazhanov,Jimbo2} (see also \cite{jimbo} and reference therein) identifying the physical model (if any) behind such a solution could be not an easy task.

\section{The Super Sinh Gordon and its non relativistic limit}
\label{SShGsec}

The SShG model, besides being integrable, it is also supersymmetric. Nowadays SUSY is a well-established subject of theoretical physics, originally born in the framework of high energy physics in $3+1$ dimensions \cite{CM1967,HLS}. However its presence is now widely spread to many different fields. For example, in the context of $1+1$ dimensional systems, it appears as a key tool to identify interesting classes of universality \cite{FQS,Q1986,ZamLG} as well as to constraint scattering amplitudes in IQFT out of criticality \cite{Sc1990} or to avoid confinement of topological excitations \cite{SUSYkink}. 

On a general ground, supersymmetric models possess an equal number of bosonic and fermionic excitations: supersymmetry exchanges ones with the others. SUSY, as other continuous symmetries, can be described as a sequence of infinitesimal transformations induced by the proper generators, i.e. the supercharges. However, while many continuous symmetries are usually implemented by generators which satisfy a Lie Algebra (and therefore the symmetry group is a Lie Group),  
in SUSY the situation is different: the supercharges, together with the Hamiltonian and the momentum, form instead a graded Lie Algebra \cite{sohnius}, in which the role of anticommutators (rather than only commutators) is fundamental.

We briefly discuss these aspects on the simple class of SUSY invariant actions which the SShG belongs to, in such a way to prepare the ground for the subsequent discussion. In particular, the SShG action is an example of a wide class of QFT with $\mathcal{N}=1$ supersymmetry \cite{mirrorbook} (i.e. with a pair of conserved supercharges)
\be
\mathcal{S}_{\text{SUSY}}=\int dx dt\;\left[\frac{1}{2}\partial_{\mu}\phi\partial^{\mu}\phi +\frac{1}{2}\bar{\chi}i\slashed{\partial}\chi-\frac{1}{2}(W'[\phi])^{2}-\frac{1}{2}\bar{\chi}\chi W''[\phi]\right]
\label{SShG5}\,\,\, ,
\ee
where $\phi$ is a bosonic real field and $\chi$ is majorana spinor which, with the following choice of the $\gamma$ matrices 
\be
\gamma^0=c\begin{pmatrix} 0&&-i\\i&&0\end{pmatrix},\hspace{2pc}\gamma^1=\begin{pmatrix} 0&&i\\i&&0\end{pmatrix},\hspace{2pc}\bar{\chi}=\chi^\dagger c^{-1}\gamma^0\,\,\, .
\ee
can be chosen to have real components 
\be
\chi=\begin{pmatrix} \chi^{+}\\ \chi^{-} \end{pmatrix},\hspace{2pc} \chi^{\pm} \text{ real fields}\,\,\,.
\ee
In Eq.\,(\ref{SShG5}) $W[\phi]$ is an analytic potential. It is immediate to see that, if in the above action we choose $W[\phi]=\frac{4mc}{g^2}\cosh(g\phi/2)$, the SShG action (\ref{1}) is readily recovered. In the perspective of taking the NR limit, we restore $c$ in the definition of the Minkowski metric
\be
\eta^{\mu\nu}=\begin{pmatrix} c^{-2} && 0\\ 0&&-1\end{pmatrix} \,\,\, .
\ee
Supersymmetric actions, even more general than the one given above and with larger supersymmetries, can be efficiently generated with the method of  superfields. An extensive discussion of the beautiful and elegant superspace formalism is beyond the purpose of the present work, that is focused on the integrability aspects rather than on the supersymmetric ones. However, the interested reader can refer to \cite{mirrorbook,superspace} for a clear and complete discussion of the subject.

Actions in the form (\ref{SShG5}) have two conserved Hermitian supercharges \cite{OW1978}, which are linear in the fermionic fields 
\begin{equation}
Q_{\pm}=\frac{1}{c}\int dx \hspace{3pt}\left(\frac{1}{c}\partial_{t}\phi \pm\partial_{x}\phi\right)\chi^{\mp}\mp W'(\phi)\chi^{\pm}\,\,\, .\label{SShG8}
\end{equation}
Their relevant commutators/anticommutators are easily computed
\be
\{Q_{\pm},Q_\pm\}=\frac{2}{c}H\pm 2P,\hspace{2pc} \{Q_{+},Q_{-}\}=0, \hspace{2pc}[H,Q_\pm]=0,\hspace{2pc}[P,Q_{\pm}]=0\,\,\, , \label{9}
\ee
where $H$ and $P$ are, respectively, the Hamiltonian and the momentum generator, i.e. $H$, $P$, $Q_\pm$ form the $\mathcal{N}=1$ superalgebra with zero central charge \cite{mirrorbook}. An additional comment is due: the absence of the central charge, i.e. the fact the $\{Q_+,Q_-\}$ vanishes, is true if the potential $W[\phi]$ does not sustain solitonic excitations. In fact, if topologically charged excitations are instead allowed, the aforementioned anticommutator does not vanish and it is equal to the topological charge \cite{OW1978}. Here, since we are ultimately interested in the SShG model, we are in the simpler situation in which topological excitations are absent, thus we have to consider the algebra given in Eq.\,(\ref{9}). 

SUSY can be equivalently implemented in terms of infinitesimal transformations generated by the supercharges
\be
\delta_\pm \phi=[Q_{\pm},\phi]=-i \chi^{\mp},\hspace{2pc}\delta_\pm\chi^\mp=\{Q_{\pm},\chi^{\mp}\}=\frac{1}{c}\partial_{t}\phi\pm\partial_{x}\phi,\hspace{2pc}\delta_\pm \chi^\pm=\{Q_{\pm},\chi^{\pm}\}=\mp W'[\phi]\, \, \,, 
\ee
and it is easy to check that such an infinitesimal transformation is indeed a symmetry for the action, $\delta S_\text{SUSY}=0$.

The presence of SUSY in the SShG plays a key role in the determination of the exact scattering matrix of the system, since this new symmetry provides extra constraints on the scattering matrix, that must be added to those coming from the unitarity and Yang Baxter conditions. In the SShG model bound states and topological excitations are absent and the particle content of the model is the same as the one guessed from perturbation theory, i.e. a bosonic and fermionic species of degenerated mass. Of course, perturbative effects dress the bare mass that appears in the action, but supersymmetry ensures the equality of the two renormalized masses. Consider now a bosonic and a fermionic single particle states $\ket{b(\theta)}$ and $\ket{f(\theta)}$, the action of the supercharges $Q_{\pm}$ on these states can be readily worked out imposing they send a boson into a fermion 
\be
Q_{\pm}\ket{b(\theta)}\propto \ket{f(\theta)},\hspace{3pc} Q_{\pm}\ket{f(\theta)}\propto \ket{b(\theta)} \,\,\, ,
\ee
while the correct prefactors follow as soon as we impose the superalgebra (\ref{9}), leading to \cite{Sc1990}
\begin{equation}
Q_{+}\ket{b(\theta)}=e^{i\pi/4} \sqrt{Mc}\;e^{\theta/2}\ket{f(\theta)},\hspace{2pc}Q_{-}\ket{b(\theta)}=ie^{i\pi/4}\sqrt{Mc}\;e^{-\theta/2}\ket{f(\theta)}\,\,\, ,\label{17}
\end{equation}
\begin{equation}
Q_{+}\ket{f(\theta)}=e^{-i\pi/4} \sqrt{Mc}\;e^{\theta/2}\ket{b(\theta)},\hspace{2pc}Q_{-}\ket{f(\theta)}=-ie^{-i\pi/4}\sqrt{Mc}\;e^{-\theta/2}\ket{b(\theta)}\,\,\, ,\label{18}
\end{equation}
where $M$ is the renormalized mass of the model. It must be said that in the above transformations there is a certain amount of freedom, since we can redefine the phases in the fermionic and bosonic states as we wish: the choice of $e^{i\pi/4}$ has been made to match the usual mode splitting of perturbation theory.
The action of the supercharges over one-particle states naturally generalizes to multiparticle states: since the charges are local, they act additively on the multiparticle states. However, because of their fermionic nature, we get an extra minus sign each time we move across a fermionic particle
\be
Q_{\pm}^{(n)}=Q_{\pm}^{(1)}\otimes 1\otimes 1\otimes ...+Q_{F}\otimes Q_{\pm}^{(1)}\otimes 1\otimes...+Q_{F}\otimes Q_{F}\otimes Q_{\pm}^{(1)}\otimes ...\,\,\, ,\label{19}
\ee
where $Q_\pm^{(n)}$ is the supercharge acting on the $n$ particles Hilbert space (obtained as tensor product of the one particle spaces), $Q_{\pm}^{(1)}$ is the supercharge acting on the one particle states and
\be
Q_F\ket{b(\theta)}=\ket{b(\theta)},\hspace{4pc}Q_F\ket{f(\theta)}=-\ket{f(\theta)}
\ee
is the fermion parity operator. 

Because of supersymmetry, the scattering matrix is strongly constrained \cite{Sc1990}. In particular the scattering matrix of SShG has been computed in \cite{A1991,A1994} and, choosing as basis $\{\ket{bb},\ket{ff},\ket{bf},\ket{fb}\}$ where $b$ is the boson and $f$ the fermion, it assumes this form:
\be
S_{\text{SShG}}(\theta)=Y(\theta)\begin{pmatrix}1-\frac{2i \sin\alpha \pi}{\sinh\theta}&& -\frac{i\sin\alpha\pi}{\cosh\frac{\theta}{2}} && 0&& 0\\ -\frac{i\sin\alpha\pi}{\cosh\frac{\theta}{2}}&& -1-\frac{2i\sin\alpha\pi}{\sinh\theta} && 0 &&0 \\ 0 && 0&& -\frac{i\sin\alpha\pi}{\sinh\frac{\theta}{2}} &&1\\ 0 && 0 &&1&& -\frac{i\sin\alpha\pi}{\sinh\frac{\theta}{2}}\end{pmatrix}\,\,\, ,\label{21}
\ee

\be
Y(\theta)=\frac{\sinh(\theta/2)}{\sinh(\theta/2)+i\sin(\alpha \pi)}\exp\left[i\int_{0}^{\infty} \frac{dt}{t} \frac{\sinh(t(1-\alpha))\sinh(t\alpha)}{\cosh t \cosh^2(t/2)}\sin(t\theta/\pi)\right]\,\,\, .\label{22}
\ee
The parameter $\alpha$ is linked to the bare coupling of the action as
\be
\alpha=\frac{ cg^2}{16\pi+ c g^2}\,\,\, .
\ee
The scattering matrix exhausts the basic informations we need in order to take the NR limit of SShG, but before doing so we should at least mention the generalities of its NR counterpart, presented in Eq.~(\ref{2}). As we already mentioned, this NR model is integrable and its particle content consists in a bosonic species and a fermionic one, as it is clear from the Hamiltonian: since the interaction is repulsive, bound states are absent. Besides being integrable, the model is also supersymmetric. However, the supersymmetry in this NR model is slightly different from its relativistic counterpart: with the same notation used in (\ref{2}) we define 
the non-hermitian NR supercharge $q$
\be
q=\int dx \, \Psi^\dagger\varphi\,\,\, .\label{24}
\ee
As it is clear, $q$ is a fermionic operation that simply replaces a boson with a fermion at the same position, the Hermitian conjugate $q^\dagger$ does the opposite operation. A simple calculation shows that both $q$ and $q^\dagger$ are indeed symmetries for the Hamiltonian $H^\text{NR}_\text{SShG}$. Unlike the relativistic case, the algebra of $q,q^\dagger$ no longer involves the Hamiltonian and the momentum, but rather the total number of particles
\be
\{q,q\}=0,\hspace{3pc}\{q^\dagger,q\}=\int dx\, \left(\Psi^\dagger\Psi+\varphi^\dagger \varphi\right)\,\,\, .
\ee
On the one particle asymptotic states the action of the supercharge is far easy, because if a  boson is present it replaces it with a fermion, otherwise it annihilates the state
\be
q\ket{b(k)}=\ket{f(k)},\hspace{4pc}q\ket{f(k)}=0\,\,\, ,\label{26}
\ee
while $q^\dagger$  exchanges the role of the fermion and of the boson. On multiparticle states, the non relativistic supercharges $q$ and $q^\dagger$ act as their relativistic cousins, i.e. as (\ref{19}). The two-body scattering matrix of the NR model is far simpler to be derived with respect to the SShG case, since the two-body eigenfunctions (and even the many-body one through the coordinate Bethe Ansatz) can be explicitly found \cite{LY1973,ID2006} and from them we read the scattering matrix. In the same basis we used for the relativistic scattering matrix, but using the momenta instead of the rapidities, it states

\be
S(k)=\begin{pmatrix}\frac{k-i2m\lambda}{k+i2m\lambda}&&0&&0&&0\\0&&-1&&0&& 0\\0&&0&&\frac{-i2m\lambda}{k+i2m\lambda}&&k\\0&&0&&k&&\frac{-i2m\lambda}{k+i2m\lambda}\\ \end{pmatrix}\, .\label{27}
\ee

\subsection{Taking the NR limit}

After this basic introduction to the SShG model and to its NR counterpart, we proceed now in taking the NR limit and show that one model is sent into the other. The identification is twofold: we firstly analyze the NR limit of the exact scattering matrix and recognize the latter to match with that of the NR model (\ref{2}). 
This simple calculation indicates the correct path for obtaining the limit: while $c$ is sent to infinite we should simultaneously tune the coupling $g$ in such a way that $gc$ remains constant in the NR limit. Not surprisingly, this is the same double-limit necessary in the Sinh Gordon case \cite{KMT,first_paper}: 
with a precise knowledge of the relevant scaling for the NR limit, we will then complete the mapping studying the dynamics from 
the action or, rather, from the equation of motion. Consider then the aforementioned NR limit on the scattering matrix (\ref{21}): using that in the NR limit 
$\theta\simeq\frac{k}{Mc}$ we readily get
\be
\lim_{\text{NR}}
S_{\text{SShG}}(\theta)=\begin{pmatrix}\frac{k-2i M\frac{c^2g^2 }{16}}{k+i2M \frac{g^2c^2}{16}}&& 0 && 0&& 0\\ 0&& -1 && 0 &&0 \\ 0 && 0&& -\frac{i2M\frac{c^2g^2}{16}}{k+i2M \frac{g^2c^2}{16}} &&\frac{k}{k+i2M \frac{g^2c^2}{16}}\\ 0 && 0 &&\frac{k}{k+i2M \frac{g^2c^2}{16}}&& -\frac{i2M\frac{c^2g^2}{16}}{k+i2M \frac{g^2c^2}{16}} \end{pmatrix}\,\,\, ,\label{28}
\ee
which coincides with (\ref{27}), provided we choose $\lambda=\frac{g^2c^2}{16}$. Besides the scattering matrix, we can readily get the NR limit of the supercharges analyzing their action on the one-particle states (\ref{17}), (\ref{18}) and (\ref{26}). In particular, we can identify
\be
\lim_{\text{NR}}\frac{e^{-i\pi/4}Q_{+}+i e^{-i\pi/4}Q_{-}}{2\sqrt{Mc}}=q\,\,\, .
\ee
Of course, besides the limit of the scattering matrix, understanding the mapping at the level of the dynamics is highly desirable. The first studies on NR limits considered the mapping at the level of the Lagrangian \cite{KMT}, but as pointed out in \cite{first_paper}, such an operation could hide non trivial subtleties. 
The most direct way to get the NR limit is to proceed from the (normal ordered) equations of motion of the model \cite{first_paper}, but the preliminary step is to correctly identify the NR fields starting from the relativistic ones. Hence, following \cite{first_paper,KMT}, we split the bosonic field in positive and negative frequencies
\be
\phi(t,x)=\frac{1}{\sqrt{2m}}\left(e^{imc^2t}\varphi^\dagger(t,x)+e^{-imc^2t}\varphi(t,x)\right)\,\,\, .\label{30}
\ee
Under the assumption of a slow dynamics for $\varphi$ in the NR limit, it can be shown that the two fields $\varphi$ and $\varphi^\dagger$ satisfy the NR canonical commutation rules \cite{first_paper,KMT}. An analogue operation must be done for the fermionic fields, thus we consider the following mode splitting \cite{dint}
\begin{equation}
\chi^{+}(t,x)=-\sqrt{\frac{ c}{2}}\left(e^{imc^2t}e^{i\pi/4}\Psi^{\dagger}(t,x)+e^{-imc^2t}e^{-i\pi/4}\Psi(t,x)\right)\,\,\, ,\label{31}
\end{equation}
\begin{equation}
\chi^{-}(t,x)=i\sqrt{\frac{ c}{2}}\left(e^{imc^2t}e^{i\pi/4}\Psi^{\dagger}(t,x)-e^{-imc^2t}e^{-i\pi/4}\Psi(t,x)\right)\,\,\, .\label{32}
\end{equation}
It is an immediate check to verify that the canonical anti-commutation rules for the Majorana fermions
\begin{equation}
\{\chi^{\pm}(t,x),\chi^{\pm}(t,y)\}= c \hspace{2pt}\delta(x-y),\hspace{4pc}\{\chi^{\pm}(t,x),\chi^{\mp}(t,y)\}=0\, \, \, ,
\end{equation}
ensure that $\Psi$ and $\Psi^\dagger$ obeys the NR anticommutation rules.

In order to extract the NR dynamics of the $\varphi$ and $\Psi$ fields we proceed through the Heisenberg equations, normal ordered with respect to the NR fields (that are directly associated to the mode expansion of the free part of the relativistic action)
\begin{equation}
\left(i\gamma^{\nu}\partial_{\nu}-mc\,:\cosh(g\phi/2):\,\right)\chi=0\,\,\, ,\label{34}
\end{equation}
\begin{equation}
\partial_\mu\partial^\mu\phi=-:\left(\frac{mc}{g}\right)^2g\sinh(g\phi):-:\frac{g mc}{4}\sinh(g\phi/2)\bar{\chi}\chi:\,\,\, .
\end{equation}
As a matter of fact, rather than Eq.\,(\ref{34}), it is much more convenient to apply $\left(-i\gamma^{\mu}\partial_{\mu}-mc\cosh(\beta\phi/2)\right)$ and reduce it to a second order differential equation
\be
\left(\partial_{\mu}\partial^{\mu}+m^2c^2:\cosh^2(g\phi/2):+:i\frac{mcg}{2}\sinh(g\phi/2)\gamma^{\mu}\partial_{\mu}\phi:\right)\chi=0 \,\,\,.
\ee
This has the nice property of being diagonal in the fermionic field. We can now easily take the NR limit plugging in the above equations the mode expansion, then taking $c\to\infty$, $g\to 0$ while $gc$ is kept constant. For $g\to 0$ the potentials in the equation of motion can be Taylor expanded: the oscillating terms in (\ref{30}), (\ref{31}), (\ref{32}) take care of the $m^2c^2$ divergent terms that appear Taylor expanding the above equations of motion. After this cancellation, we are left with finite quantities and oscillating terms. Matching then the fast oscillating phases finally leads to the desired NR equation of motion. 
A careful discussion of the limit procedure can be found in \cite{first_paper}; here we directly report the result of this lengthy but straightforward calculation 
\be
i\partial_{t}\Psi^{\dagger}= \frac{\partial_{x}^{2}\Psi^{\dagger}}{2m}-\frac{g^2c^2}{8} \varphi^{\dagger}\varphi \Psi^{\dagger},\hspace{3pc}
i\partial_{t}\varphi^{\dagger}=\frac{\partial_{x}^2\varphi^{\dagger}}{2m} - \frac{g^2c^2}{8}\varphi^{\dagger}\varphi^{\dagger}\varphi-\frac{g^2c^2}{8}\varphi^{\dagger}\Psi^{\dagger}\Psi\,\,\, .
\ee
These equations of motion are exactly those that can be derived from the NR Hamiltonian (\ref{2}), provided we identify $\lambda=\frac{g^2c^2}{16}$, consistently with what we already knew from the scattering matrix (in the latter, we also need $\lim_\text{NR}M=m$). With the NR limit of the dynamics, we conclude the identification of (\ref{2}) as the proper NR limit of the SShG model.

\section{The Gross-Neveu and its non relativistic limit}
\label{GNsec}

After the SShG model, the second example we are interested in is the Gross-Neveu model. Its action is given by 
\be
\mathcal{S}_{\text{GN}}=\int dxdt\; \left[\sum_{j=1}^N\frac{1}{2}\bar{\chi}_{j}i\slashed{\partial}\chi_j+\frac{\beta}{N}\left(\sum_{j=1}^N\bar{\chi}_{j}\chi_j\right)^2 \right]\, \, \, 
\label{GNagain}
\ee
and represents a purely fermionic theory based on $N$ Majorana fermions with a $O(N)$ symmetric coupling. Despite the absence of an explicit mass term in the action (\ref{GNagain}), a mass is dynamically generated and therefore the fermions $\chi_i$ are effectively massive \cite{GN1978}. For this reason, the use of na\"ive perturbation theory is problematic. In order to control the qualitative and quantitative features of the GN model is instead better to use the \emph{large $N$ expansion} techniques \cite{MZ2003}, by means of which we are able to recognise GN as an asymptotically free theory with a certain number of bound states \cite{GN1978,DHA1975}. In addition to these features, the GN model has also been identified to be integrable \cite{ZZ1978}, a crucial information which leads to the computation of the scattering matrix of the elementary particles \cite{ZZ}
\be
S_{ab}^{cd}(\theta)=\delta_a^b\delta_c^d S^{(1)}(\theta)+\delta_a^d\delta_b^c S^{(2)}(\theta) +\delta_a^c\delta_b^d S^{(3)}(\theta)\,\,\, ,
\ee
where the amplitudes
\be
S^{(3)}(\theta)=-\frac{1}{N-2}\frac{i2\pi}{\theta}S^{2}(\theta),\hspace{3pc}S^{(1)}(\theta)=-\frac{1}{N-2}\frac{i2\pi}{i\pi-\theta}S^{2}(\theta),\hspace{2pc}
S^{(2)}(\theta)=U(\theta)U(i\pi-\theta)\,\,\, 
\ee
are parameterised in terms of $\Gamma$ functions
\be
U(\theta)=\frac{\Gamma\left(-\frac{1}{N-2}-i\frac{\theta}{2\pi}\right)\Gamma\left(\frac{1}{2}-i\frac{\theta}{2\pi}\right)}{\Gamma\left(\frac{1}{2}-\frac{1}{N-2}-i\frac{\theta}{2\pi}\right)\Gamma\left(-i\frac{\theta}{2\pi}\right)}\,\,\, .
\ee
As for the SShG case, the scattering matrix is the starting point to understand the NR limit of the model. Note that the scattering amplitudes above are completely independent from the coupling constant $\beta$ of the GN model: as a matter of fact, due to the mass transmutation, the coupling of the four-fermions interaction in the GN action enters in all the physical quantities simply through the renormalised mass $m$, that can be equivalently used to parameterise the theory \cite{ZZ,GN1978,DHA1975}. The n\"aive NR limit would imply a vanishing rapidity $\theta\simeq \frac{k}{mc}$ and leads to a trivial result. Thus, we are forced to change the only parameter that enters in the scattering amplitudes, namely the number of particle species $N$. Indeed, it is easy to check that a sensible result for the scattering matrix can be achieved taking the double limit $c,N\to \infty$, but keeping $Nc^{-1}$ constant
\be
\lim_\text{NR}S_{ab}^{cd}(\theta)=\delta_a^d\delta_b^c \frac{k}{k-i2\pi mcN^{-1}} +\delta_a^c\delta_b^d \frac{-i2\pi mcN^{-1}}{k-i2\pi mcN^{-1}}\,\,\, ,
\ee
This scattering matrix is in perfect agreement with that of the candidate NR model in Eq.~(\ref{4}), provided we choose $\lambda=-\pi cN^{-1}$, that is kept constant in the NR limit. Note that the presence of bound states in the GN model is reflected in the attractive interaction in (\ref{4}). Beside the check on the scattering matrixes, to accomplish the NR limit we must study the dynamics as well. However the procedure to tackle this problem is more involved than the SShG case: 
indeed the necessity of taking $N\to\infty$, as well as the non perturbative nature of the theory, makes impossible to proceed directly from the equation of motion and therefore we have to look for other methods. The key observation is the form of the coupling of the NR model: large values of $Nc^{-1}$ correspond to the small coupling regime of the NR model, therefore, we can reasonably expect that, in the NR limit, the large $N$ expansion of the GN model will reproduce the perturbative expansion of the NR model. Indeed, this is the same scheme encountered in extracting the NR limit of the non-linear sigma model \cite{first_paper}: in fact, in this section (and in dealing with the SNLS), we closely follow the same steps used and explained in \cite{first_paper}. The NR limit is based on the same mode-splitting we adopted for the SShG model in eqs.~(\ref{31}, \ref{32}). The strategy also consists of to go to the Fourier space and recognise that the correlators of $\Psi^\dagger_j$ and $\Psi_j$ on the ground state of the GN model reduce to the NR correlators on the NR vacuum (i.e. the state with no particles). This identification will be carried on at the level of Feynman graphs.

As we already commented in Section \ref{introGN}, the $N\to \infty$ limit leads to an infinite number of fermionic species in the NR model. However, notice that the Hamiltonian in Eq.~(\ref{4}) conserves separately the number of each species. Therefore, we are allowed to compute correlation functions of the NR model 
(\ref{4}) where the internal index $j$ runs only from $1$ to $n$ {\em finite} different species, even though $N\to\infty$: in this way, the the NR limit of the GN model reproduces simultaneously all the models (\ref{4}) with an arbitrary number of species.

\subsection{The large $N$ expansion of the Gross-Neveu model}
\label{GNNsec}

As we just commented, the starting point to understand the NR limit of the GN model is its large $N$ expansion, thus it is our intention to provide a short summary of this technique. The procedure starts with a formal manipulation of the action (\ref{3}) by he insertion of an auxiliary (or ghost) field
\be
\mathcal{S}_\text{GN}^\text{aux}=\int dxdt\; \left[\sum_{j}\frac{1}{2}\bar{\chi}_{j}i\slashed{\partial}\chi_j-\frac{N}{16}\beta^{-1}\Lambda^2-\frac{1}{2}\Lambda \left(\sum_{j}\bar{\chi}_{j}\chi_j\right)\right]\,\,\, .
\ee
Recovering the GN action from the above is immediate: the ghost field does not propagate and performing the gaussian path integral in 
$\Lambda$ simply amounts to replace $\Lambda\to \frac{4\beta}{N}\left(\sum\bar{\chi}_j\chi_j\right)$. However, a judicious use of the ghost field naturally leads to the large $N$ expansion we are looking for.  As a preliminary step, since we know in advance that the fermions dynamically acquire a mass, we shift the ghost field 
$\Lambda\to \Lambda+mc$ (this operation is allowed by the fact that $\Lambda$ is a dummy integration variable in the path integral)
\be
\mathcal{S}_\text{GN}^\text{aux}=\int dxdt\; \left[\sum_{j}\frac{1}{2}\bar{\chi}_{j}\left(i\slashed{\partial}-mc\right)\chi_j-\frac{N}{16}\beta^{-1}(\Lambda+mc)^2-\frac{1}{2}\Lambda \left(\sum_{j}\bar{\chi}_{j}\chi_j\right)\right]\,\,\, .\label{44}
\ee
\begin{figure}[t]
\begin{center}
\includegraphics[scale=0.4]{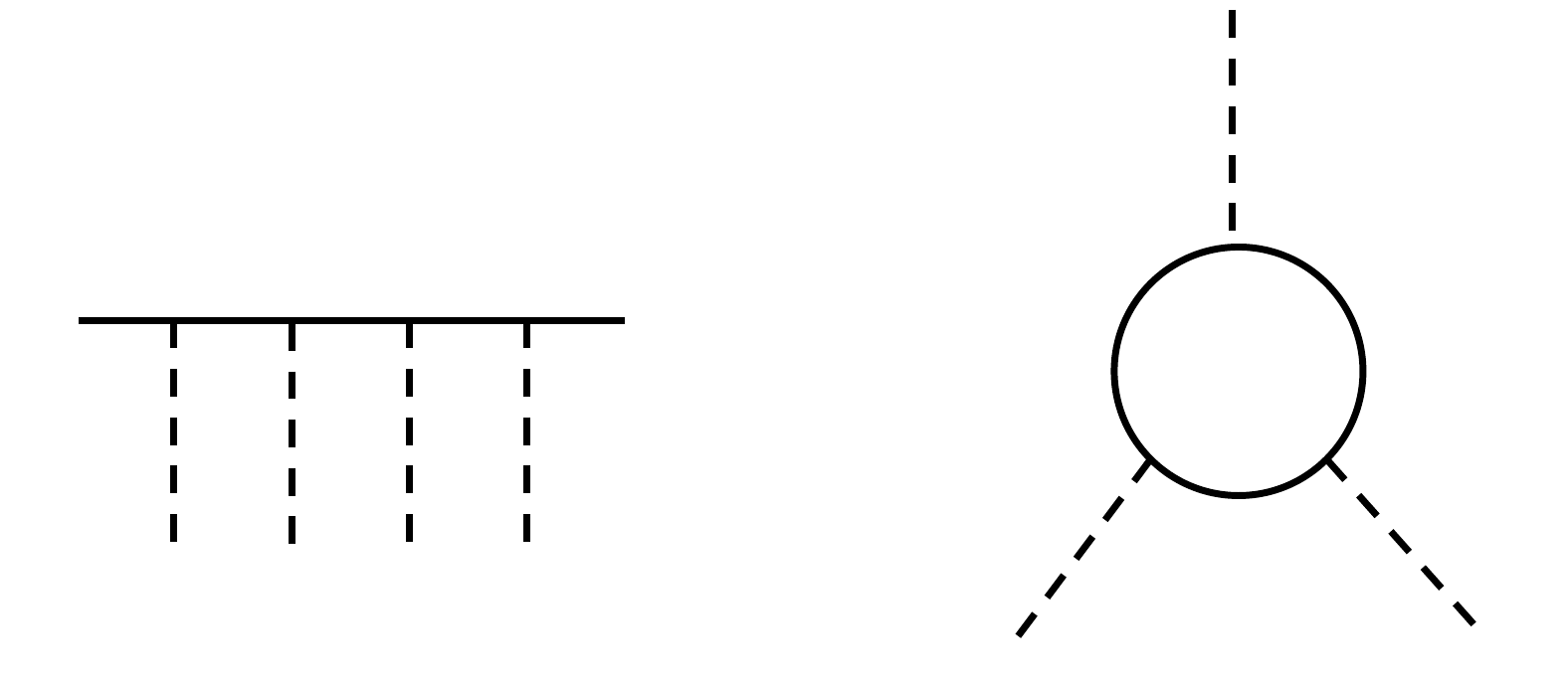}
\caption{\emph{In the Feynman diagrams that describe the path integral in $\chi_j$ at fixed ghost field there are two classes of connected diagrams. The graphs involving external continuous legs (left) and the graphs in which only the external legs of the ghost field appear (right).}}\label{fig2}
\end{center}
\end{figure}
For the time being $m$ is a free parameter that we will fix self-consistently later. As we have previously commented, performing the path integral in $\Lambda$ leads us back to the GN model, but we can equivalently integrate first over the fermionic fields and leave the ghost-field integration for a second moment. In principle, we could compute exactly the path integral in terms of the fermionic fields, but it is far more convenient to use Feynman diagrams from the beginning: 
this will automatically lead to the Feynman graphs of the large $N$ expansion.

In these Feynman graphs we then represent the free-fermionic propagators induced by the free massive part of the action (\ref{44}) as continuous lines, the ghost field is instead a dotted line and interacts with two identical fermionic fields. At this stage, we are integrating only over the fermionic fields, thus the ghost field does not propagate and behaves as an external source coupled to the fermions. To summarize, these Feynman diagrams have two kinds of external legs: continuous external legs represent the fermionic fields of the correlation function we are actually computing, then there are dotted external legs representing the ``external source" $\Lambda$. 

These considerations immediately organise the (connected) Feynman diagrams into two classes: those that possess external continuous legs and those with only dotted external legs, as described in FIG.\,\ref{fig2}. This concludes the integration over the fermions, and we now turn to the integration over the ghost field
\be
\braket{...}=\int \mathcal{D}\Lambda\; e^{i\int dxdt\; -\frac{N}{16}\beta^{-1}(\Lambda+mc)^2}\; \; \;\sum\left[\text{Feynman diagrams}\right] \; \; \; ,\label{45}
\ee
where the Feynman diagrams are of course $\Lambda$ dependent because of the dotted external legs.
\begin{figure}[t]
\begin{center}
\includegraphics[scale=0.4]{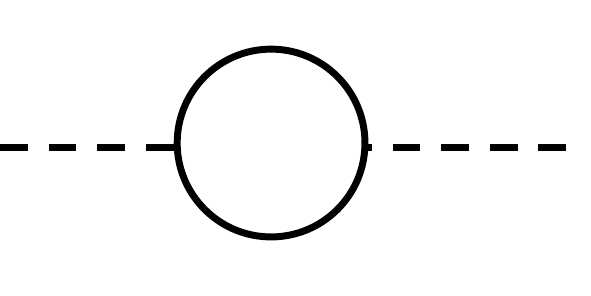}
\caption{\emph{The connected diagram that contributes to the gaussian part of the effective action in the large $N$ limit.}}\label{fig3}
\end{center}
\end{figure}
There is of course no hope in computing exactly the above path integral and we should rely on perturbation theory, starting from a gaussian action for the ghost field: a quadratic term is already evident in the above exponential, but we can do better through a partial resummation of the Feynman diagrams. In particular, among the various connected Feynman diagrams with only dotted external legs (thus associated with $\Lambda$), we consider those with two legs drawn in 
FIG.\,\ref{fig3}. This graph, having only two dotted legs, is quadratic in the ghost field: we label as $\mathcal{G}$ both this graph and its value. As it is standard in QFT \cite{weinberg}, we rewrite the sum over all diagrams with two dotted external legs as an exponential involving the sum over only the corresponding connected diagrams, i.e. $\mathcal{G}$: 
\be
\sum\left[\text{Feynman graphs}\right]=\sum\left[\text{Feynman graphs without }\mathcal{G}\right]e^{\mathcal{G}}\,\,\, . 
\ee
This allows us to rewrite Eq.~(\ref{45})  as
\be
\braket{...}=\int \mathcal{D}\Lambda\; e^{i\int dxdt\; -\frac{N}{16}\beta^{-1}(\Lambda+mc)^2+\mathcal{G}}\; \; \;\sum\left[\text{Feynman graphs without }\mathcal{G}\right]\,\,\, .\label{47}
\ee
The quadratic part of the exponential is now used to construct the propagator of the ghost field. Its exact value is computed 
in Appendix \ref{Gpropsec} and it reads
\be
G_\Lambda(k^\mu)=\left[i N\left(\frac{\beta^{-1}}{8}-\int \frac{d^2k}{(2\pi)^2}\frac{1}{k_\mu k^\mu-m^2c^2+i\epsilon}\right)-\frac{iNc}{4\pi}\sqrt{\frac{4-\frac{k^\mu k_\mu}{m^2c^2}}{-\frac{k_\mu k^\mu}{m^2c^2}}}\log\left(\frac{\sqrt{4-\frac{k_\mu k^\mu}{m^2c^2}}+\sqrt{-\frac{k_\mu k^\mu}{m^2c^2}}}{\sqrt{4-\frac{k_\mu k^\mu}{m^2c^2}}-\sqrt{-\frac{k_\mu k^\mu}{m^2c^2}}}\right)\right]^{-1}\, .\label{48}
\ee
Note that the integral in the Eq.~\eqref{48} is an UV divergent constant: this can be in principle absorbed in a renormalization of the coupling $\beta$, but this coupling does not appear only in the propagator, since in the exponential of (\ref{47}) it is also associated with a linear term in $\Lambda$ obtained from the expansion of the square $(\Lambda + mc)^2$. We postpone the issue of the renormalization of $\beta$ to the discussion of Feynman rules and why these provide the sought large $N$ expansion. Now that the ghost field has a proper propagator, the final Feynman diagrams can be obtained from the previous ones simply letting the dotted lines propagate and using Eq.~(\ref{48}) for each dotted propagator. The field $\Lambda$ interacts with two fermionic fields as it happened in the previous Feynman diagrams, but an extra interaction vertex involving only one dotted line results from the linear term in the exponential of (\ref{47}).

In the simple framework we just described, an extra selection rule must be kept in mind: in Eq.\,(\ref{47}) we removed the graph $\mathcal{G}$ from the Feynman diagrams and used it in the definition of the propagator $G_\Lambda$. Because of this reason, the $\mathcal{G}$ graph cannot appear in the final Feynman diagrams, even though it is an internal graph: we forcefully remove all the diagrams that contain an internal loop with only two departing dotted lines.
Finally, it should be stressed that $\Lambda$ is a ghost field, i.e. it never enters in correlation functions: at the level of Feynman diagrams, this simply means that dotted lines are now always internal propagators and never external legs.

\begin{figure}[b]
\begin{center}
\includegraphics[scale=0.3]{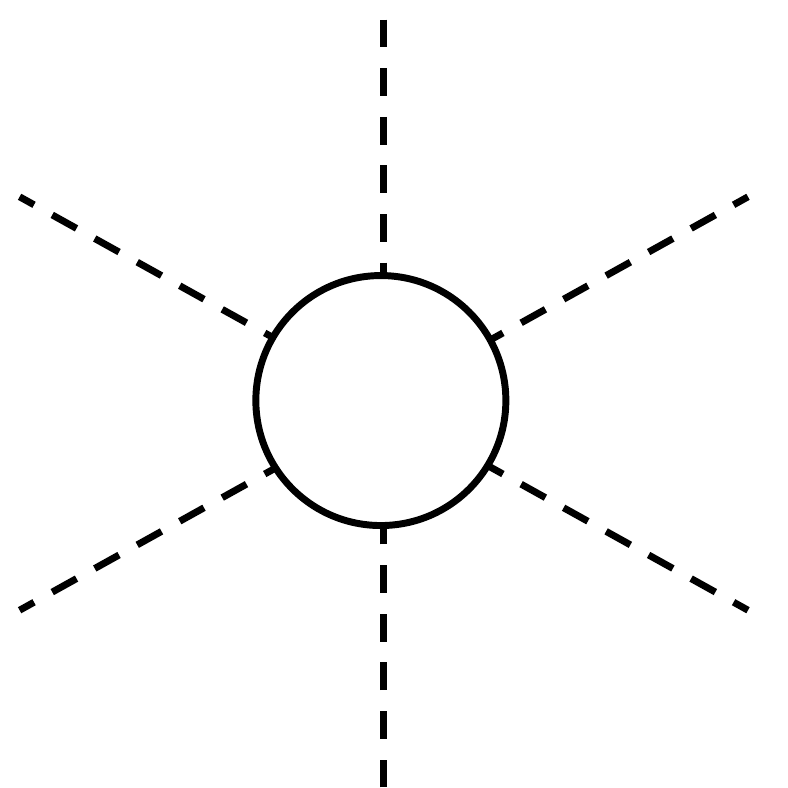}
\caption{\emph{The integration over the internal loops of continuous lines gives a $N$ factor.}}\label{fig4}
\end{center}
\end{figure}

We are now ready to discuss why these Feynman rules provide the large $N$ expansion, by mean of a simple power counting of the $N$ factors in each diagram.
First of all, note that the $G_\Lambda$ propagator carries a factor $1/N$, while the interaction between the ghost and the fermions does not have any explicit $N$ power, thus it is $\mathcal{O}(N^0)$. It follows that $N$ divergent contributions can only come from internal continuous loops as shown in FIG.\,\ref{fig4}: since internal loops imply a summation over the $N$ internal fermionic degrees of freedom, each of these loops is proportional to $N$.

In order to understand how the contribution of propagators can balance the $N$ divergence of the loops, it is useful to split the $N^{-1}$ factor of the $G_\Lambda$ propagator as $N^{-1}=(1/\sqrt{N})(1/\sqrt{N})$ and attach one factor $1/\sqrt{N}$ at each edge of the dotted lines.
\begin{figure}[t]
\begin{center}
\includegraphics[scale=0.3]{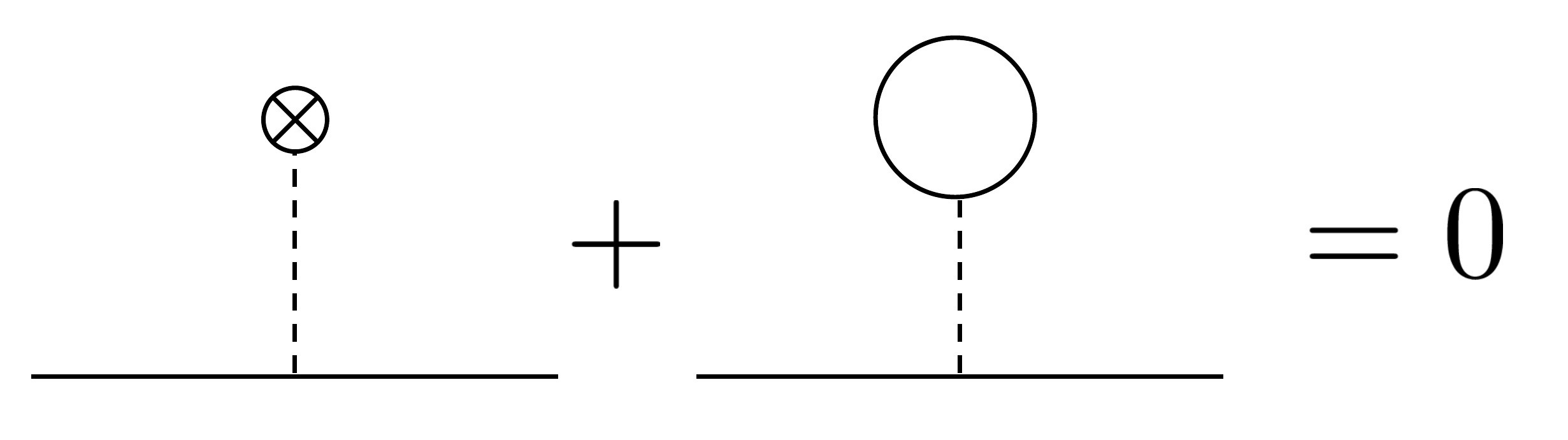}
\caption{\emph{Cancellation of $N$ divergent graphs. The crossed vertex represents the interaction vertex associated with the $\Lambda-$linear term in Eq. (\ref{47}).}}\label{fig5}
\end{center}
\end{figure}
\begin{figure}[b]
\begin{center}
\includegraphics[scale=0.35]{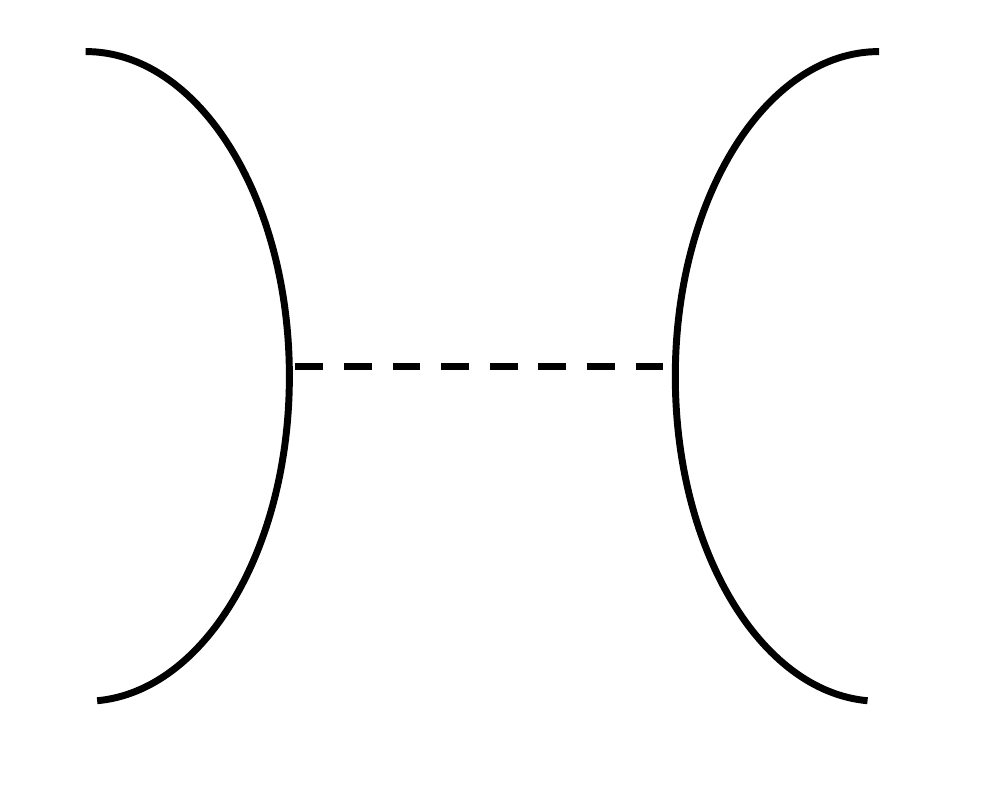}
\caption{\emph{Four field vertex mediated by the ghost field.}}\label{fig6}
\end{center}
\end{figure}
With this convention, an internal loop with $n$ departing dotted lines is now associated with a $N^{1-n/2}$ factor, thus we learn that all the loops with $n\ge 3$ departing dotted lines are indeed suppressed in the large $N$ limit. Since, as we commented before, internal loops with two departing dotted legs are forcefully removed from the Feynman diagrams, the only problematic loops are those with a single departing dotted leg. However, we can now take advantage of the floating mass $m$ and remove the problematic loops cancelling them with the $\Lambda-$linear term of (\ref{47}), as shown in FIG.\,\ref{fig5}. This request is translated into a mass equation that links together the coupling $\beta^{-1}$ with $m$
\be
\frac{\beta^{-1}}{8}-\int \frac{d^2k}{(2\pi)^2}\frac{1}{k_\mu k^\mu-m^2c^2+i\epsilon}=0\,\,\, .\label{49}
\ee
Note that the above UV divergent integral is the same it appears in the $G_\Lambda$ propagator (\ref{48}), thus the mass equation also cures the UV divergence of the propagator that simply becomes
\be
G_\Lambda(k^\mu)=-\frac{i4\pi}{Nc}\sqrt{\frac{-\frac{k_\mu k^\mu}{m^2c^2}}{4-\frac{k^\mu k_\mu}{m^2c^2}}}\frac{1}{\log\left(\frac{\sqrt{4-\frac{k_\mu k^\mu}{m^2c^2}}+\sqrt{-\frac{k_\mu k^\mu}{m^2c^2}}}{\sqrt{4-\frac{k_\mu k^\mu}{m^2c^2}}-\sqrt{-\frac{k_\mu k^\mu}{m^2c^2}}}\right)}\,\,\, .\label{53}
\ee
The mass equation also makes clear the aforementioned mass transmutation: after we impose eq.\,(\ref{49}), the effect of $\beta$ on the Feynman diagrams appears only through the mass $m$. Thus, we can use the mass parameter $m$ rather than $\beta$ as the free parameter and regard instead $\beta$ as a function of the mass.

Let's close this section with the observation that all this construction can be compactly encoded in an effective action, with some extra caveat we are going to describe. The key observation is that, since dotted lines appear only as internal propagators, they are always embedded in the graph of FIG.\,\ref{fig6} , that can be thus regarded as an effective four fermions interaction mediated by the ghost propagator. Pursuing this interpretation, we can write an effective action where only the fermionic fields appear
\be
\mathcal{S}_\text{eff}=\mathcal{S}_\text{free}+\int d^2x d^2y\,\, \sum_{j,j'}\bar{\chi}_j(x^\mu)\chi_{j}(x^\mu)\frac{iG_\Lambda(x^\mu-y^\mu)}{8}\bar{\chi}_{j'}(y^\mu)\chi_{j'}(y^\mu)\,\,\, ,\label{51}
\ee
Where $\mathcal{S}_\text{free}$ is the free action of the massive fermions and with $G_\Lambda(x^\mu-y^\mu)$ we simply mean the ghost propagator in the coordinate space. However, the correspondence between the above effective action and the Feynman diagrams of the large $N$ expansion is not perfect and must be corrected with the following selection rules:
\begin{enumerate}
\item The mass equation cancels closed loops with only a departing dotted leg (FIG.\,\ref{fig5}). This rule is not already implemented in the effective action, however the aforementioned loops can be removed by a simple mass counterterm (i.e. the net effect of a mass counterterm is the same as the crossed vertex of FIG. \ref{fig5}).
\item The large $N$ expansion prevents the appearance of closed loops which would correspond to FIG.\,\ref{fig3}. As a matter of fact, these loops were used to define the $G_\Lambda$ propagator, thus their presence is already kept in account in the effective interaction vertex and they must be removed by hand from the Feynman diagrams.
\end{enumerate}
It must be said that the simple framework discussed so far deals with the bare theory but divergent Feynman diagrams can be encountered: these needs a proper renormalization \cite{abdalla,polyakov}. However, such a procedure reduces to the renormalization of the mass (or equivalently a renormalization of $\beta$, 
thanks to the mass transmutation) and renormalization of the field strength, which eventually amounts to include a $\sum_j\bar{\chi}_j\slashed{\partial}\chi_j$ counterterm in the action.

\subsection{Non relativistic limit from the large $N$ expansion}
\label{GNNRsec}
We are now ready to extract the non relativistic limit of the GN model from its large $N$ expansion. The starting point is still the same mode splitting of the Majorana fermions introduced in Eqs.~(\ref{31}, \ref{32}) during the analysis of the SShG model. However, we look at it in the Fourier space, 
where the oscillating phases means translations of the momentum along the time direction
\begin{subequations}
\begin{align}
&\Psi_j(k^0,k^1)=-\frac{e^{i\pi/4}}{\sqrt{2c}}\left(\chi_j^+(k^0-mc^2,k^1)+i\chi_j^{-}(k^0-mc^2,k^1)\right)\,\,\, ,\label{55}\\
&\Psi_j^\dagger(k^0,k^1)=-\frac{e^{-i\pi/4}}{\sqrt{2c}}\left(\chi_j^+(-k^0+mc^2,-k^1)-i\chi_j^{-}(-k^0+mc^2,-k^1)\right)\,\,\, .\label{56}
\end{align}
\end{subequations}
As a first step, we show that the free propagator for the $\Psi_j$ fields reduces, in the NR limit, to the proper NR propagator. 
Using the above relations and the free propagator for the Majorana fermions we get
\be
\braket{\Psi_{j}^\dagger(k^0,k^1)\Psi_{j'}(q^0,q^1)}_\text{free}=(2\pi)^2\delta_{jj'}\delta(k^\mu-q^\mu)\left[\frac{i\frac{k_0}{c^2}-i2m}{\frac{(-k_0+mc^2)^2}{c^2}-k_1^2-m^2c^2+i\epsilon}\right]\,\,\, .
\ee
Using the fact that in the above is meant $\epsilon\to 0^+$, we can equivalently write (except for 
an inessential redefinition of $\epsilon$): 
\begin{eqnarray}
\nonumber&&\braket{\Psi_{j}^\dagger(k^0,k^1)\Psi_{j'}(q^0,q^1)}_\text{free}=(2\pi)^2\delta_{jj'}\delta(k^\mu-q^\mu)
\frac{2mc\left(1-\frac{k_0}{2mc^2}\right)}{2\sqrt{k_1^2+m^2c^2}}\times\\
&&\times \left[\frac{-i}{-k_0+mc^2-c\sqrt{k_1^2+m^2c^2}+i\epsilon}-\frac{-i}{-k_0+mc^2+c\sqrt{k_1^2+m^2c^2}-i\epsilon}\right]\,\,\, .
\end{eqnarray}
Instead the $\Psi^\dagger\Psi^\dagger$ propagator is:
\be
\braket{\Psi_j^\dagger(k^0,k^1)\Psi_{j'}^{\dagger}(q^0,q^1)}_0=(2\pi)^2\delta_{jj'}\delta(k^\mu+q^\mu-2mc^2)\frac{\frac{k_1}{c}}{\frac{(-k_0+mc^2)^2}{c^2}-k_1^2-m^2c^2+i\epsilon}\,\,\,.
\ee
The NR limit of the propagators ($c\to\infty$ at fixed momenta) is now immediate and we recover the retarded free propagator of non relativistic fermions
\be
\lim_{\text{NR}}\braket{\Psi_j^\dagger(q^0,q^1)\Psi_{j'}(k^0,k^1)}_0=-(2\pi)^2\delta_{jj'}\delta(k^\mu-q^\mu)\frac{-i}{k_0+\frac{k_1^2}{2m}-i\epsilon},\hspace{2pc}
\lim_{\text{NR}}\braket{\Psi_j^\dagger(k^0,k^1)\Psi_{j'}^{\dagger}(q^0,q^1)}_0=0\,\,\, .\label{57}
\ee
Therefore, the limit over the free propagators is correct. Let's turn to the discussion about the effect of the interactions. As a preliminary step, let's write the effective action in terms of the $\Psi_j$ fields in the coordinate space thanks to the simple observation
\be
\frac{1}{2}\bar{\chi}_j(x^\mu)\chi_j(x^\mu)=c\,\Psi^\dagger(x^\mu)\Psi(x^\mu)\,\,\, ,\label{58}
\ee
as it is immediately derived from the mode expansion in Eqs.~(\ref{31}, \ref{32}) and the fact that $\Psi^\dagger, \Psi$ are grassmanian fields. Therefore we have 
\be
\mathcal{S}_\text{eff}=\mathcal{S}_\text{free}+\int d^2x d^2y\,\, \sum_{j,j'}\Psi^\dagger_j(x^\mu)\Psi_{j}(x^\mu)\frac{ic^2G_\Lambda(x^\mu-y^\mu)}{2}\Psi^\dagger_{j'}(y^\mu)\Psi_{j'}(y^\mu)\,\,\, .
\ee
In order to study the NR limit, we rewrite the above in momentum space
\be
\mathcal{S}_\text{GN}^\text{eff}=\mathcal{S}_\text{free}+\int \frac{d^8k}{(2\pi)^8}(2\pi)^2\delta\left(k_1^\mu+k_3^\mu-k_2^\mu-k_4^\mu\right)\,\, \sum_{j,j'}\Psi^\dagger_j(k_1^\mu)\Psi_{j}(k_2^\mu)\frac{ic^2G_\Lambda(k_1^\mu+k_2^\mu)}{2}\Psi^\dagger_{j'}(k_3^\mu)\Psi_{j'}(k_4^\mu)\,\,\, .
\ee
Note that the structure of the above interaction already reproduces the one of the putative NR model in eq.\,(\ref{4}). From this expression, the non-relativistic limit of the \emph{tree-level} Feynman diagrams is immediately recovered. First of all, the additional selection rules necessary to extract the large $N$ expansion from the effective action are not necessary at tree level, since they only concern loops. Moreover, in absence of momentum integrations (required for loops), 
the limit of the whole Feynman diagram simply coincides with the limit of its elementary components, i.e. propagators and interaction vertices. Since in the NR limit the $\Psi\Psi$ propagators vanish, the conservation laws ensure that all momenta in the internal propagators (and therefore in the interaction vertices) 
attain their non-relativistic values.

At the end of this discussion, the NR limit of the tree-level Feynman graphs are the diagrams constructed with the NR propagator 
(\ref{57}) and the NR limit of the interaction vertex
\be
\lim_\text{NR}\frac{ic^2G_\Lambda(k_1^\mu+k_2^\mu)}{2}=\pi cN^{-1}\,\,\, ,
\ee
which coincides with the coupling obtained from the scattering matrix. The reader could be worried by an apparent sign discrepancy from the above and the coupling obtained from the scattering matrix, that requires $\lambda=-\pi cN^{-1}$ in the Hamiltonian (\ref{4}). However there is no inconsistency at all, 
since the Feynman diagrams in the NR model are computed from the Lagrangian rather than the Hamiltonian: in the Lagrangian the potential energy appears with the reversed sign and this gives the extra minus we are apparently missing.

With this last calculation the NR limit of the dynamics at three level is complete and indeed we recover the correspondence between GN and its proposed NR limit.
We think that the limit of the scattering matrix together with this tree-level calculation provides sufficient indications to establish the Hamiltonian in (\ref{4}) as the NR limit of the GN model beyond any reasonable doubt. However, we give further comments about what happens beyond the tree-level calculation.

\subsubsection{A glimpse beyond tree level \label{glimpsetree}}

\begin{figure}[t]
\begin{center}
\includegraphics[scale=0.3]{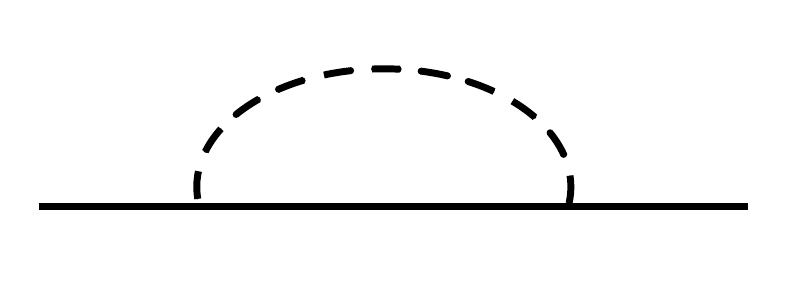}
\caption{\emph{Graph removed by normal ordering.}}\label{fig7}
\end{center}
\end{figure}
Beyond tree level, some technical issues complicate the NR limit. Actually, these apparent problems are analogous to those encountered in \cite{first_paper} while studying the NR limit of the non-linear sigma model, which also requires a large-$N$ expansion. However, as we already stressed, the issues that we are going to present are expected to not spoil the NR limit, since the latter is unequivocally identified by the scattering matrix and the tree level computation. In fact, in \cite{first_paper} the corrections beyond tree level have been tested in the non-linear sigma model and the NR limit was not ruined, as expected. A similar check can be performed even in the GN case leading to identical conclusions. To keep the discussion at the simplest possible level, we limit ourselves to describe the various issues and how they are successfully addressed.
\begin{enumerate}
\item The first difficulty (absent at tree level) is that the relativistic effective vertex that appears in the effective action (\ref{51}) is not normal ordered, differently from the NR vertex. At the level of Feynman graphs, the normal ordering recipe prevents the legs of the vertex of FIG.\,\ref{fig6} to self-interact: 
while one self-interaction has already been eliminated by the mass equation (FIG.\,\ref{fig5}), we are still left with the ``unwanted" graph of FIG.\,\ref{fig7}, which is instead absent in the NR model. Since this graph has two legs, we could ask ourselves if it can be naturally canceled by a field strength and a mass renormalisation. While such a cancellation is not expected to be possible in the relativistic theory because of the non-trivial momentum dependence of the graph, we expect however that such a cancellation occurs in the NR limit: this has been explicitly checked in \cite{first_paper} in the analogous case of the non-linear sigma model.

\item On the large-$N$ diagrams we should enforce the extra selection rule that prevents the presence of internal loops such as FIG.\,\ref{3}. This rule seems instead absent in the NR model. However, although not explicit, such a rule is naturally implemented in the NR diagrams because of time ordering. To understand this more in detail, consider the internal loops of FIG.\,\ref{fig3}, obtained patching together two interaction vertexes
\be
\int d^2x d^2 y\, \lambda^2\sum_{j,j',l,l'} \Psi^\dagger_{j'}(x^\mu)\Psi_{j'}(x^\mu)\braket{\Psi^\dagger_j(x^\mu)\Psi_l(y^\mu)}_\text{free}\braket{\Psi^\dagger_{l}(y^\mu)\Psi_{j}(x^\mu)}_\text{free}\Psi^\dagger_{j'}(y^\mu)\Psi_{j'}(y^\mu)\,\,\, .\label{61}
\ee
Here we assume that  the uncontracted fields will be contracted to others interaction vertexes in the Feynman diagrams, or end as external legs. The loop itself is simply the contraction
\be
\braket{\Psi^\dagger_j(x^\mu)\Psi_l(y^\mu)}_\text{free}\braket{\Psi^\dagger_{l}(y^\mu)\Psi_{j}(x^\mu)}_\text{free}\,\,\, .
\ee
However, the propagators are actually \emph{time-ordered propagators}, therefore $\braket{\Psi^\dagger_j(x^\mu)\Psi_l(y^\mu)}_\text{free}$ is different from zero only when $y^0>x^0$, while $\braket{\Psi^\dagger_{l}(y^\mu)\Psi_{j}(x^\mu)}_\text{free}$ does not vanish only if $x^0>y^0$: therefore, the NR loop (\ref{61}) always vanishes.

\item The NR limit needs to be taken on the full Feynman diagrams and not on their building blocks, i.e. interaction vertexes and propagators. While the two limits coincide at tree level, this is no longer true as soon as we need to integrate over internal momenta: in particular, the relativistic theory presents UV divergences that must be renormalized before of taking the NR limit. 
\end{enumerate}

In order to check that the aforementioned issues do not spoil the NR limit, in \cite{first_paper} the limit of the vectorial $\mathcal{O}(N)$ non-linear sigma model has been performed over the $\mathcal{O}(N^{-1})$ correlation functions (that in the NR model corresponds to the first order in the coupling), where a non-trivial renormalization occurs. In particular, it appears the analogue of the graph of FIG.\,\ref{fig7}: its UV divergence was absorbed in a redefinition of the mass equation and the finite part disappeared as it should.
A similar computation can be done also for the GN model though involving lengthier calculations; we choose to not report them for simplicity.

\section{The supersymmetric non-linear sigma model and its non-relativistic limit}
\label{SNLS}

The SNLS model, whose action is 
\be
\mathcal{S}_\text{SNLS}=\int dxdt\;\left[\sum_j \frac{1}{2}\partial_\mu\phi_j\partial^\mu\phi_j +\frac{1}{2}\sum_j\bar{\chi}_ji\slashed{\partial}\chi_j+\frac{\beta}{N}\left(\sum_j\bar{\chi}_j\chi_j\right)^2\right]\,\,\, ,\label{ns55}
\ee
subjected to the constraints
\be
\sum_j\phi_j^2=\frac{N}{8}\beta^{-1}\,\,\, ,\hspace{3pc}\sum_j \phi_j\chi_j=0\,\,\, ,\label{ns66}
\ee
can be regarded as a supersymmetric extension of the bosonic non-linear sigma model, as well as of the Gross-Neveu model. Like both of these relativistic theories, the SNLS is an integrable theory whose mass is dynamically generated by the interactions, despite the absence of an explicit mass term in the action \cite{A1978}. Similarly to both its non-supersymmetric partners, its NR limit emerges once more from its large $N$ expansion. This is indeed confirmed by the limit of the two-body scattering matrix, that is most compactly expressed in a rotated basis \cite{Witten-Shankar}: starting from the bosonic and fermionic asymtpotic states, we define the following states (without loss of generality, we can sit in the center of mass frame)
\be
\ket{S^{ab}}=\frac{1}{\sqrt{\cosh\theta}}\left[\cosh\left(\frac{\theta}{2}\right)\ket{b^a(\theta) b^b(-\theta)}+\sinh\left(\frac{\theta}{2}\right)\ket{f^a(\theta)f^b(-\theta)}\right]\,\,\, ,
\ee
\be
\ket{T^{ab}}=\frac{1}{\sqrt{\cosh\theta}}\left[-\sinh\left(\frac{\theta}{2}\right)\ket{b^a(\theta)b^b(-\theta)}+\cosh\left(\frac{\theta}{2}\right)\ket{f^a(\theta)f^b(-\theta)}\right]\,\,\, ,
\ee
\be
\ket{U^{ab}}=\frac{1}{\sqrt{2}}\left[\ket{b^a(\theta)f^b(-\theta)}+\ket{f^a(\theta)b^b(-\theta)}\right],\hspace{2pc}
\ket{V^{ab}}=\frac{1}{\sqrt{2}}\left[\ket{b^a(\theta)f^b(-\theta)}-\ket{f^a(\theta)b^b(-\theta)}\right]\,\,\, .
\ee
The SUSY invariant scattering matrix is diagonal on each set of states
\be
\braket{S^{cd}_\text{out}|S^{ab}_\text{in}}=\braket{U^{cd}_\text{out}|U^{ab}_\text{in}}=S_1(2\theta)\delta^{ac}\delta^{bd}+S_2(2\theta)\delta^{ab}\delta^{cd}+S_3(2\theta)\delta^{ad}\delta^{bc}\,\,\, ,
\ee
\be
\braket{T^{cd}_\text{out}|T^{ab}_\text{in}}=\braket{V^{cd}_\text{out}|V^{ab}_\text{in}}=\mathcal{T}_1(2\theta)\delta^{ac}\delta^{bd}+\mathcal{T}_2(2\theta)\delta^{ab}\delta^{cd}+\mathcal{T}_3(2\theta)\delta^{ad}\delta^{bc}\,\,\, ,
\ee
 with the following scattering amplitudes \cite{Witten-Shankar}
\be
S_1(\theta)=\left(1-\frac{i\sin\left(\frac{\pi}{N-2}\right)}{\sinh\left(\frac{\theta}{2}\right)}\right)S(\theta),\hspace{2pc}
S_2(\theta)=-\frac{2\pi i}{N-2}\frac{S_1(\theta)}{i\pi-\theta},\hspace{2pc}
S_3(\theta)=-\frac{2\pi i}{N-2}\frac{S_1(\theta)}{\theta}\, \, \, ,
\ee
\be
\mathcal{T}_1(\theta)=\left(1+\frac{i\sin\left(\frac{\pi}{N-2}\right)}{\sinh\left(\frac{\theta}{2}\right)}\right)S(\theta),\hspace{2pc}
\mathcal{T}_2(\theta)=-\frac{2\pi i}{N-2}\frac{\mathcal{T}_1(\theta)}{i\pi-\theta},\hspace{2pc}
\mathcal{T}_3(\theta)=-\frac{2\pi i}{N-2}\frac{\mathcal{T}_1(\theta)}{\theta}\, \, \, .
\ee
Above we have $S(\theta)=R_1(\theta)R_1(i\pi-\theta)R_2(\theta)R_2(i\pi-\theta)$ and:
\be
R_1(\theta)=\frac{\Gamma\left(\frac{1}{N-2}-\frac{i\theta}{2\pi}\right)\Gamma\left(\frac{1}{2}-\frac{i\theta}{2\pi}\right)}{\Gamma\left(-\frac{i\theta}{2\pi}\right)\Gamma\left(\frac{1}{2}+\frac{1}{N-2}-\frac{i\theta}{2\pi}\right)}\,\,\, ,
\ee

\be
R_2(\theta)=\frac{\Gamma\left(-\frac{i\theta}{2\pi}\right)}{\Gamma\left(\frac{1}{2}-\frac{i\theta}{2\pi}\right)}\prod_{l=1}^\infty\frac{\Gamma\left(\frac{1}{N-2}-\frac{i\theta}{2\pi}+l\right)\Gamma\left(-\frac{1}{N-2}-\frac{i\theta}{2\pi}+l-1\right)\Gamma^2\left(-\frac{i\theta}{2\pi}+l-\frac{1}{2}\right)}{\Gamma\left(\frac{1}{N-2}-\frac{i\theta}{2\pi}+l+\frac{1}{2}\right)\Gamma\left(-\frac{1}{N-2}-\frac{i\theta}{2\pi}+l-\frac{1}{2}\right)\Gamma^2\left(-\frac{i\theta}{2\pi}+l-1\right)}\, \, \, .
\ee
The two supercharges $Q_\pm$ act on the asymptotic states as those of the SShG, sending a fermion $f^a$ in a boson $b^a$ and vice versa: notice that supersymmetry sends $\ket{S}$ states in $\ket{U}$ states (and the opposite) and $\ket{T}$ states in $\ket{V}$ states, but the two pairs are not mixed. This is indeed the reason why the scattering amplitudes of these states are pairwise equal.

Despite the rather complicated appearance, the NR limit is rather simple to be extracted and leads to a simple result. First of all, notice that in the NR limit the rotated basis reduces to
\be
\lim_\text{NR}\ket{S^{ab}}=\ket{s^{ab}}=\ket{b^a(k)b^b(-k)},\hspace{1.5pc}\lim_\text{NR}\ket{U^{ab}}=\ket{u^{ab}}=\frac{1}{\sqrt{2}}\left[\ket{b^a(k)f^b(-k)}+\ket{f^a(k)b^b(-k)}\right]\,\,\, ,
\ee
\be
\lim_\text{NR}\ket{T^{ab}}=\ket{t^{ab}}=\ket{f^a(k)f^b(-k)},\hspace{1.5pc}\lim_\text{NR}\ket{V^{ab}}=\ket{v^{ab}}=\frac{1}{\sqrt{2}}\left[\ket{b^a(k)f^b(-k)}-\ket{f^a(k)b^b(-k)}\right]\,\,\, .
\ee
As for the GN model, a sensible limit of the scattering amplitudes is obtained letting $c\to\infty$ and $N\to\infty$, but with $N/c$ constant. In this case, the simple result is
\be
\braket{s^{cd}_\text{out}|s^{ab}_\text{in}}=\braket{u^{cd}_\text{out}|u^{ab}_\text{in}}=\frac{2k}{2k+i2\pi mcN^{-1}}\delta^{ac}\delta^{bd}-\frac{i2\pi mc N^{-1}}{2k+i2\pi mc N^{-1}}\delta^{ad}\delta^{bc}\,\,\, ,
\ee
\be
\braket{t^{cd}_\text{out}|t^{ab}_\text{in}}=\braket{v^{cd}_\text{out}|v^{ab}_\text{in}}=\frac{2k}{2k-i2\pi mcN^{-1}}\delta^{ac}\delta^{bd}-\frac{i2\pi mc N^{-1}}{2k-i2\pi mc N^{-1}}\delta^{ad}\delta^{bc}\,\,\, .
\ee
In the two lines above we can easily recognise the scattering matrix of a repulsive multicomponent Lieb-Liniger bosonic gas and that of an attractive multicomponent Lieb Liniger based on fermions \cite{yang}.

The limit of the scattering matrix matches with the qualitative prediction we can drag at first glance from the SNLS's action: being a mixture of the Gross-Neveu and the non-linear sigma model, the NR limit was expected to be a mixture of the NR limits of each model, as it can be seen from the boson-boson and fermion-fermion scattering (respectively, $\ket{s}$ and $\ket{t}$ states). However, the NR Hamiltonian associated with such a scattering matrix is not simply the sum of the bosonic and the fermionic multicomponent Lieb Liniger: bosons and fermions support a non-trivial scattering ($\ket{u}$ and $\ket{v}$ states), which must reflect on their reciprocal interaction. Based on the scattering matrix limit, we are immediately led to an Hamiltonian of the form
\be
H^\text{NR}_\text{SNLS}=\int dx \; \sum_{j}\left(\frac{\partial_x\Psi_j\partial_x\Psi_j}{2m}+\frac{\partial_x\varphi_j\partial_x\varphi_j}{2m}\right)+\lambda\sum_{j,j'}\left(\varphi_j^\dagger\varphi_{j'}^\dagger\varphi_{j'}\varphi_j-\Psi^\dagger_j\Psi^\dagger_{j'}\Psi_{j'}\Psi_j\right)+\mathcal{I}_{bf}\,\,\, ,
\ee
where the choice $\lambda=\pi cN^{-1}$ provides the matching of the boson-boson and fermion-fermion scattering with the non relativistic scattering matrix. In $\mathcal{I}_{bf}$ we encoded a yet unspecified interaction among bosons and fermions: rather than reconstructing $\mathcal{I}_{bf}$ from the scattering matrix, there is a shortcut that permits to write it down immediately, thanks to supersymmetry. Indeed, being the starting model supersymmetric, we must get a supersymmetric NR model as well, as it happened in the SShG case: asking $H^\text{NR}_\text{SNLS}$ to be symmetric under the action of the supercharges (\ref{24}) immediately leads to the Hamiltonian (\ref{n7}), i.e.
\be
H_\text{SNLS}^\text{NR}=\int dx\; \sum_j\frac{\partial_x \varphi^\dagger_j\partial_x \varphi_j}{2m}+\sum_j\frac{\partial_x \Psi^\dagger_j\partial_x \Psi_j}{2m}+\lambda\sum_{j,j'}\left(\varphi^\dagger_j\varphi^\dagger_{j'}\varphi_{j'}\varphi_j- \Psi^\dagger_j\Psi^\dagger_{j'}\Psi_{j'}\Psi_j+2 \varphi^\dagger_j\Psi^\dagger_{j'}\varphi_{j'}\Psi_j\right)\,\,\, .
\ee
It is then an easy check to show that the scattering matrix of the above matches with the NR limit previously derived: indeed, this model is a particular case of a wider class of NR supersymmetric models known to be integrable \cite{FPZ1988}.

This exhausts what can be learnt from the scattering matrix and let's now discuss the NR limit from the dynamical point of view: extracting the NR limit requires passing through the large $N$ expansion as we did in the GN case and it has been done in \cite{first_paper} for the purely bosonic non-linear sigma model. 
In light of these similarities, we avoid to repeat all the calculations of Section \ref{GNsec} and rather focus on the things which are different.

\subsection{The large $N$ expansion of SNLS and its non relativistic limit}

As we previously discussed in the case of the GN model, the large $N$ expansion of the SNLS can be derived through the insertion of suitable ghost fields. Since the action of SNLS contains the GN model we need the same ghost we used in the latter, however two additional ghosts are needed
to implement the constraints (\ref{n6}), thanks to a path integral representation of the functional Dirac $\delta$ \cite{A1978}:
\begin{subequations}
\begin{align}
&\delta\left(\sum_j\phi_j^2-\frac{N}{8}\beta^{-1}\right)=\int \mathcal{D}\Lambda' \;e^{i\int dxdt\;-\frac{\Lambda'}{2}\left(\sum_j\phi_j^2-\frac{N}{8} \beta^{-1}\right)}\,\,\, ,\\
&\delta\left(\sum_j\phi_j\chi_j^+\right)\delta\left(\sum_l\phi_j\chi_j^-\right)=\int \mathcal{D}\zeta \;e^{i\int dx dt\;  \sum_j\phi_j\bar{\zeta}\chi_j}\,\,\, .
\end{align}
\end{subequations}
Here, $\Lambda'$ is a bosonic field and $\zeta$ is instead a Majorana spinor. Thanks to these ghosts, the SNLS can be rewritten in terms of an unconstrained auxiliary action:
\begin{eqnarray}
\mathcal{S}_\text{SNLS}^\text{aux} &= & i\int dxdt\; \left[\sum_{j}\left(\frac{1}{2}\partial_\mu\phi\partial^\mu\phi+\frac{1}{2}\bar{\chi}_{j}i\slashed{\partial}\chi_j\right)-\frac{N\beta^{-1}}{16}\Lambda^2-\frac{\Lambda}{2} \left(\sum_{j}\bar{\chi}_{j}\chi_j\right) + 
\nonumber \right.\\
&  &\hspace{8mm} -\left. \frac{\Lambda'}{2} \left(\sum_j\phi_j^2-\frac{N\beta^{-1}}{8} \right) +  \sum_j\phi_j\bar{\zeta}\chi_j\right]\,\,\, .
\end{eqnarray}
Taking advantage of the fact that the interactions dynamically generate a mass for the bosons and the fermions, we shift the fields $\Lambda$ and $\Lambda'$ to make explicit a mass term in the action. Being the model supersymmetric, the mass of the fermions must be equal to the bosonic one, thus we shift
\be
\Lambda\to \Lambda+mc,\hspace{4pc} \Lambda'\to\Lambda'+m^2c^2\,\,\, .
\ee
After this shift, we can isolate a free massive action for the bosonic and fermionic fields, plus the interactions with the ghosts
\be
\mathcal{S}_\text{SNLS}^\text{aux}=\mathcal{S}_\text{free}+\int dxdt\;\left[ -\frac{N\beta^{-1}}{16}(\Lambda+mc)^2-\frac{\Lambda}{2} \left(\sum_{j}\bar{\chi}_{j}\chi_j\right)-\frac{\Lambda'}{2}\left(\sum_j\phi_j^2-\frac{N\beta^{-1}}{8} \right)+ \sum_j\phi_j\bar{\zeta}\chi_j \right]\,\,\, .
\ee
From this action with the ghosts fields, through the same passages we followed in the GN case, it is possible to define an effective action without ghosts
\begin{eqnarray}
\nonumber&&\mathcal{S}_\text{SNLS}^\text{eff}=\mathcal{S}_\text{free}+\int d^2x d^2y\;\left[\frac{i}{8}\sum_{jj'} \bar{\chi}_j(x^\mu)\chi_j(x^\mu) \bar{\chi}_{j'}(y^\mu)\chi_{j'}(y^\mu)G_\Lambda(x^\mu-y^\mu)+\right.\\
&&\left.+\frac{i}{2}\sum_{j,j'}\bar{\chi}_j(x^\mu)G_\zeta(x^\mu-y^\mu)\chi_{j'}(y^\mu)\phi_j(x^\mu)\phi_{j}(y^\mu)+\frac{i}{8} \sum_{jj'}\phi_j^2(x^\mu)\phi_{j'}^2(y^\mu)G_{\Lambda'}(x^\mu-y^\mu)\right]\,\,\, ,\label{87}
\end{eqnarray}
that naturally embeds the large $N$ expansion of the model, provided we forcefully implement the same selection rules we discussed at the ends of Section \ref{GNNsec}. The propagator of the $\Lambda$ ghost $G_\Lambda$ is exactly the same of that obtained in the Gross-Neveu model in (\ref{53}).

Similarly, the $\Lambda'$ propagator coincides with that obtained in the purely bosonic non-linear sigma model \cite{first_paper}, i.e. (in the Fourier space)
\be
G_{\Lambda'}(k^\mu)=i2\pi cN^{-1} m^2 \left(4-\frac{k_\mu k^\mu}{m^2c^2}\right)\sqrt{\frac{-\frac{k_\mu k^\mu}{m^2c^2}}{4-\frac{k_\mu k^\mu}{m^2c^2}}}\frac{2}{\log\left(\frac{\sqrt{4-\frac{k_\mu k^\mu}{m^2c^2}}+\sqrt{-\frac{k_\mu k^\mu}{m^2c^2}}}{\sqrt{4-\frac{k_\mu k^\mu}{m^2c^2}}-\sqrt{-\frac{k_\mu k^\mu}{m^2c^2}}}\right)}\,\,\, .\label{88}
\ee
Finally, the $G_\zeta(k^\mu)$ propagator has the form
\be
G_{\zeta}(k^\mu)=-i\frac{2\pi cN^{-1}}{c^2}\left[k_\mu\gamma^\mu-2mc\right]\sqrt{\frac{-\frac{k_\mu k^\mu}{m^2c^2}}{4-\frac{k_\mu k^\mu}{m^2c^2}}}\frac{2}{\log\left(\frac{\sqrt{4-\frac{k_\mu k^\mu}{m^2c^2}}+\sqrt{-\frac{k_\mu k^\mu}{m^2c^2}}}{\sqrt{4-\frac{k_\mu k^\mu}{m^2c^2}}-\sqrt{-\frac{k_\mu k^\mu}{m^2c^2}}}\right)}\,\,\, .
\ee
The calculation of these propagators closely resembles that of the $G_\Lambda$ propagator we report in Appendix \ref{Gpropsec}, substituting bosonic propagators with fermionic ones when needed.

One may wonder what happens to the mass equation, that has not been mentioned yet. Following the same steps of the GN case, in the SNLS we would reach two mass equations to cancel the $N$ divergent loops (FIG. \ref{fig5}): in one case the dotted line is the $\Lambda$ propagator and the closed loop is made of fermions, in the other case the dotted propagator corresponds to $\Lambda'$ and the closed loop is constructed out of bosons. Thus, we get respectively a mass equation for the fermions and one for the bosons: in a generic model this would have implied a different mass for fermions and bosons, but this is not the case for a supersymmetric theory as the SNLS that forcefully possesses mass degeneracy.
Indeed, a direct computation of the two mass equations shows they are perfectly equivalent and thus we consistently have the same mass for the fermionic and the bosonic particles, being fixed by the same mass equation of the GN model (\ref{49}).

Armed with the effective action (\ref{87}), we can now take the NR limit of the model: for simplicity, we focus only on tree level Feynman diagrams, keeping in mind that beyond the tree level, we face the same problems described in the GN case in Section \ref{glimpsetree}. The first step is, as usual, the mode splitting of the relativistic fields and, as in the GN case, we need it in the Fourier space: regarding the fermions, we can indeed use again (\ref{55}-\ref{56}). In the case of the bosonic fields we use the same mode splitting introduced in \cite{first_paper}, i.e.
\be
\varphi_{j}(k^0,k^1)=\sqrt{2m}\;\phi_{j}(k^0-mc^2,k^1)\;\Theta(mc^2-k^0),\hspace{2pc}
\varphi_{j}^{\dagger}(-k^0,-k^1)=\sqrt{2m}\;\phi_{j}(k^0+mc^2,k^1)\;\Theta(mc^2+k^0)\,\,\, ,\label{90}
\ee
whose consistency with the splitting in the coordinate space (\ref{30}) is immediate. It is not difficult to show, by mean of the same passages of the fermionic case of Section \ref{GNNRsec}, that the propagators of the so defined $\varphi_j$ fields reduce, in the NR limit, to the NR retarded free propagator of bosonic particles: the reader can refer to \cite{first_paper} for the explicit check.

We can now consider the interaction as we did in the GN model: in the NR limit of the tree-level Feynman graphs the relevant interactions are only those vertexes in which \emph{all} the non-relativistic fields attain non relativistic momenta. However, while this only possible choice for the four-fermions interaction, 
this is not true for the other two cases: we recall the discussion for the four-bosons interaction already presented in \cite{first_paper} and then discuss the mixed term.

Expressing the aforementioned interaction term in the Fourier space and substituting the relativistic fields with the mode splitting (\ref{90}), we discover 
that there are two inequivalent terms in which all the NR fields can attain non-relativistic moments at the same time, consistently with the global momentum conservation. These terms are
\begin{align}
&\int \frac{d^8k}{(2\pi)^8}(2\pi)^2\delta\left(k^\mu_1+k^\mu_2-k^\mu_3-k^\mu_4\right)\;\frac{i}{2(2m)^2} \sum_{jj'}\varphi_j^\dagger(k^\mu_1)\varphi^\dagger_{j'}(k_2^\mu)\varphi_j(k_3^\mu)\varphi_{j'}(k_4^\mu)\;G_{\Lambda'}(k_1^\mu-k_3^\mu)\,\,\, \\
&\int \frac{d^8k}{(2\pi)^8}(2\pi)^2\delta\left(k^\mu_1+k^\mu_2-k^\mu_3-k^\mu_4\right)\;\frac{i}{4(2m)^2} \sum_{jj'}\varphi_j^\dagger(k^\mu_1)\varphi^\dagger_j(k_2^\mu)\varphi_{j'}(k_3^\mu)\varphi_{j'}(k_4^\mu)\;G_{\Lambda'}(k_2^\mu+k_3^\mu+2mc^2\delta_0^\mu)\,\,\, .
\end{align}
While the index structure of the first term is the same as the NR four-bosons interaction, the second term does not have an analogous one in the NR model. 
Thus, in order to have that the NR limit of the tree-level diagrams matches the NR model in Eq.~(\ref{n7}), we need
\be
\lim_\text{NR}\frac{i}{2(2m)^2}G_{\Lambda'}(k_1^\mu-k_3^\mu)=-\pi cN^{-1},\hspace{4pc}\lim_\text{NR}\frac{i}{4(2m)^2}G_{\Lambda'}(k_2^\mu+k_3^\mu+2mc^2\delta_0^\mu)=0\,\,\, .
\ee
This is indeed the case as it can be easily verified from the explicit expression (\ref{88}). A similar analysis should be now repeated for the fermions-bosons interaction: going in Fourier transform and plugging the mode splitting, we find again two contributions that are relevant in the NR limit, however only the term with the correct indexes structure survives in the end. We skip here these calculations because they resemble those for the four-bosons terms we just discussed: 
the NR limit of the fermions-bosons vertex exactly matches the corresponding non-relativistic interaction vertex in the Hamiltonian (\ref{n7}), as it should be.
This check concludes the NR mapping of the dynamics between the SNLS and the NRIM (\ref{n7}), at least at the tree level: however, being both integrable models and given the non-perturbative check over the scattering matrices, it is hard to believe that calculations beyond tree level could spoil the aforementioned identification.

\section{Conclusions}
\label{conclusions}

In this work we extend to the fermionic case the analysis previously done of the non relativistic integrable models which emerge as proper limits of relativistic integrable field theories. In particular, we were interested to understanding which kind of non-relativistic integrable models are obtained in the presence of fermions.  The analysis carried here on three different significant models seems once again to confirm the presence of a tight bottleneck of models when the non-relativistic limit is taken: namely, relativistic theories that may greatly differ from each others in the non-relativistic limit are nevertheless always projected on Lieb Liniger like models. 

From a qualitative point of view, it is the locality of the original interactions of the relativistic models that is responsible for the  ubiquitous presence of Lieb-Liniger like models, as discussed here after. Let the dynamics of the NRIM be governed in fact by an Hamiltonian $H=T+V$, where $T$ is the kinetic part and $V$ the proper interaction: of course, the kinetic part obtained in the NR limit is forced to be that of non-relativistic particles. Considering the interaction term we expect it to be: 
\begin{itemize}
\item ultra-local, i.e. made of an integral over the space of powers of the fields and their derivatives at the same point, since this was the form of the original relativistic interaction;
\item an interaction which respects the conservation of the number of particles, for its integrable nature; 
\item 
an interaction which acts pairwise on the particles, since integrability requires the factorisation of the multi-particle scattering in two-body processes. 
\end{itemize}
All these considerations lead to the following generic form of the interaction
\be
V=\int dx \sum_{a,b,c,d}\Lambda^{a,b,c,d}\, \psi_a^\dagger\psi_b^\dagger\psi_c\psi_d\, ,
\ee
where the fields can be either bosons of fermions, $\Lambda^{a,b,c,d}$ is a coupling constant. 
In principle, there is not any obstruction to the presence of higher derivative terms in the interaction, as it happens in the NR limit of the Thirring model \cite{dint}, however this is not the case for the models analyzed in the present work.
When a mass degeneracy is present, the tunneling process among particles of different species is admissible: this indeed happens in the SNLS model where bosons and fermions can exchange their internal degrees of freedom, while this feature was absent in the cases analyzed in \cite{first_paper}. Even though the constraints on the interaction are based on heuristic considerations, the aforementioned picture is perfectly respected by the models analysed in this work and in \cite{first_paper} and, on the basis of this analysis, one can also expect the other known relativistic models to be not an exception.

\,\,\, \emph{Acknowledgements:} 
We would like to thank Fabian Essler for pointing out ref. \cite{FPZ1988}, we acknowledge one of the referees for driving our attention towards Ref. \cite{dint,dint2}. A.B. thanks the Rudolf Peierls Centre for 
Theoretical Physics of Oxford for the ospitality during the last part of this work. A. D. L. thanks the EPSRC Quantum Matter in and out of Equilibrium Ref. EP/N01930X/1.

\newpage 

\appendix

\section{The ghost propagator}
\label{Gpropsec}

The purpose of this section is providing the analytical computation of the ghost propagator $G_\Lambda$ (\ref{48}), used both in Section \ref{GNsec} and Section \ref{SNLS}. As explained in Section \ref{GNNsec}, this propagator is constructed out of the explicit $\Lambda^2$ term in (\ref{47}) plus the contribution of the loop diagram of FIG. \ref{fig3}. In particular, we have
\be
G_\Lambda^{-1}(k^\mu)=\frac{iN}{8}\beta^{-1}-\frac{N}{2}\int \frac{d^2 q}{(2\pi)^2}\text{Tr}\left[\frac{q^\mu\gamma_\mu+mc}{q^\mu q_\mu-m^2c^2+i\epsilon_1}\frac{\left(q^\mu-k^\mu\right)\gamma_\mu+mc}{\left(q^\mu-k^\mu\right)\left(q_\mu-k_\mu\right)-m^2c^2+i\epsilon_2}\right]\,\,\, .
\ee

As first step, we compute the trace over the $\gamma$ matrices
\be
G_\Lambda^{-1}(k^\mu)=\frac{iN}{8}\beta^{-1}-N\int \frac{d^2 q}{(2\pi)^2}\frac{q^\mu(q_\mu-k_\mu)+m^2c^2}{q^\mu q_\mu-m^2c^2+i\epsilon_1}\frac{1}{\left(q^\mu-k^\mu\right)\left(q_\mu-k_\mu\right)-m^2c^2+i\epsilon_2}\,\,\, .
\ee

First of all, it must be noticed that the above integral is UV divergent, as it is clear from the momenta power counting.. However, its divergence can be readily isolated as follows
\begin{eqnarray}
\nonumber&&\int \frac{d^2 q}{(2\pi)^2}\frac{q^\mu(q_\mu-k_\mu)+m^2c^2}{q^\mu q_\mu-m^2c^2+i\epsilon_1}\frac{1}{\left(q^\mu-k^\mu\right)\left(q_\mu-k_\mu\right)-m^2c^2+i\epsilon_2}=\\
\nonumber&&=\int \frac{d^2 q}{(2\pi)^2}\frac{1}{q^\mu q_\mu-m^2c^2+i\epsilon_1}\left(1+\frac{q^\mu k_\mu+2m^2c^2-k_\mu k^\mu}{\left(q^\mu-k^\mu\right)\left(q_\mu-k_\mu\right)-m^2c^2+i\epsilon_2}\right)=\\
\nonumber&&=\int \frac{d^2 q}{(2\pi)^2}\frac{1}{q^\mu q_\mu-m^2c^2+i\epsilon_1}+\int \frac{d^2 q}{(2\pi)^2}\frac{q^\mu k_\mu+2m^2c^2-k_\mu k^\mu}{q^\mu q_\mu-m^2c^2+i\epsilon_1}\frac{1}{\left(k^\mu-q^\mu\right)\left(k_\mu-q_\mu\right)-m^2c^2+i\epsilon_2}\,\,\, .\\
\end{eqnarray}

Above, the first integral is a UV divergent constant (i.e. independent from the injected momentum $k$) while the second integral is convergent and we are going to compute it.
The first step is to combine a Wick rotation with a rescaling of the variables in such a way the Minkowski metric is replaced with the standard euclidean metric. The substitution amounts to replace $(q^0,q^1)=(imc^2 x^0,mcx^1)$, consistently we define the euclidean vector $s$ in such a way $(k^0,k^1)=(imc^2s^0,mcs^1)$, in this way
\begin{eqnarray}
\nonumber&&\int \frac{d^2 q}{(2\pi)^2}\frac{ q^\mu k_\mu+2m^2c^2-k_\mu k^\mu}{q^\mu q_\mu-m^2c^2+i\epsilon_1}\frac{1}{\left(k^\mu-q^\mu\right)\left(k_\mu-q_\mu\right)-m^2c^2+i\epsilon_2}\hspace{1pc}\longrightarrow\\
&&\longrightarrow\hspace{1pc} ic\int \frac{d^2 x}{(2\pi)^2}\frac{\textbf{s}^2-\textbf{s}\textbf{x}+2}{\textbf{x}^2+1}\frac{1}{(\textbf{x}-\textbf{s})(\textbf{x}-\textbf{s})+1}\,\,\, .
\end{eqnarray}
The last integral can be computed rewriting the integrand by mean of the Feynman's trick
\be
\frac{1}{AB}=\int_0^1d\xi\;\frac{1}{\left[\xi A+(1-\xi)B\right]^2}\,\,\, ,
\ee
that permits to reduce to a spherically symmetric integral
\be
ic\int_0^1d\xi \int \frac{d^2 x}{(2\pi)^2}\frac{\textbf{s}^2-\textbf{s}\textbf{x}+2}{\left[\xi(\textbf{x}^2+1)+(1-\xi)[(\textbf{x}-\textbf{s})(\textbf{x}-\textbf{s})+1]\right]^2}=ic\int_0^1d\xi \int \frac{d^2 x}{(2\pi)^2}\frac{\xi \textbf{s}^2-\textbf{s}\textbf{x}+2}{\left[\textbf{x}^2+(1-\xi)\xi\textbf{s}^2+1\right]^2}
\ee

The last integration is easily performed, as well as the integration over the auxiliary parameter $\xi$, giving
\be
\frac{ic}{4\pi}\sqrt{\frac{4+\textbf{s}^2}{\textbf{s}^2}}\log\left[\frac{\sqrt{4+\textbf{s}^2}+\sqrt{\textbf{s}^2}}{\sqrt{4+\textbf{s}^2}-\sqrt{\textbf{s}^2}}\right]\,\,\, .
\ee
The last passage is to rewrite this expression in terms of the $k^\mu$ momentum $\textbf{s}^2=-\frac{k_\mu k^\mu}{m^2c^2}$. Thus, we finally get

\be
G_\Lambda(k^\mu)=\left[i N\left(\frac{\beta^{-1}}{8}-\int \frac{d^2k}{(2\pi)^2}\frac{1}{k_\mu k^\mu-m^2c^2+i\epsilon}\right)-\frac{iNc}{4\pi}\sqrt{\frac{4-\frac{k^\mu k_\mu}{m^2c^2}}{-\frac{k_\mu k^\mu}{m^2c^2}}}\log\left(\frac{\sqrt{4-\frac{k_\mu k^\mu}{m^2c^2}}+\sqrt{-\frac{k_\mu k^\mu}{m^2c^2}}}{\sqrt{4-\frac{k_\mu k^\mu}{m^2c^2}}-\sqrt{-\frac{k_\mu k^\mu}{m^2c^2}}}\right)\right]^{-1}\,\,\, .
\ee

\end{document}